\documentstyle[eqsecnum,preprint,aps]{revtex}
\tighten
\draft
\begin{document}
\widetext
\preprint{CLNS 93/1243}
\bigskip
\bigskip
\title{Kac and New Determinants for\\
Fractional Superconformal Algebras}
\medskip
\author{Zurab Kakushadze\cite{foot1} and S.-H. Henry Tye}
\bigskip
\address{Newman Laboratory of Nuclear Studies,
Cornell University,
Ithaca, NY 14853-5001, USA}
\date{October 1993}
\bigskip
\medskip
\maketitle
\begin{abstract}

We derive the Kac and new
determinant formulae for an arbitrary (integer) level $K$ fractional
superconformal
algebra using the BRST cohomology techniques developed in conformal
field theory. In particular, we reproduce the Kac determinants for the
Virasoro ($K=1$) and superconformal ($K=2$) algebras. For $K\geq3$
there always exist modules
where the Kac determinant factorizes into a product of more fundamental new
determinants. Using our results
for general $K$, we
sketch the non-unitarity proof for
the $SU(2)$ minimal series; as expected, the only unitary models are those
already known from the coset construction. We apply the Kac determinant
formulae for the spin-4/3 parafermion current algebra ({\em i.e.}, the
$K=4$ fractional superconformal algebra) to the recently
constructed three-dimensional flat Minkowski space-time representation of the
spin-4/3 fractional superstring. We prove the no-ghost theorem for the
space-time bosonic sector of this theory; that is, its physical spectrum
is free of negative-norm states.
\end{abstract}

%============================================================================
\widetext
\newpage
\section{Introduction}
\bigskip

{}The Kac determinant \cite{Kac} for the Virasoro algebra
is a very useful tool for analyzing the minimal unitary
series in CFT \cite{FQS1} and for understanding
the no-ghost theorem in the bosonic string
theory \cite{Thorn}.
So far, the Kac determinants have been known only for the
Virasoro and superconformal algebras \cite{FQS2}. These algebras
are special cases of the so-called fractional superconformal algebras (FSCAs)
\cite{open,AGT} labeled by the level $K$ of the $SU(2)_K$
current algebra \cite{KZ} (The level $K=1$ and $2$ FSCAs are precisely
the Virasoro and superconformal algebras).
In this paper we derive the Kac and new determinant formulae
for an arbitrary (integer) level $K$ FSCA.

{}In section II we classify the modules of the level $K$ FSCA and
present the Kac determinant formulae for each module. For $K\geq3$ there
always exist
modules where the Kac determinant factorizes into a product of more
fundamental new determinants. For these modules we also present the new
determinant formulae.
We recover the well-known Kac determinant formulae for the Virasoro
and superconformal algebras, and elaborate on the case of
the spin-4/3 parafermion current algebra ({\em i.e.}, the $K=4$
FSCA) of Zamolodchikov and Fateev \cite{FZft}. This algebra has three
modules (the so-called $S$-, $D$- and $R$-modules) as opposed to the
superconformal algebra that has only two modules ({\em i.e.}, the
Neveu-Schwarz and Ramond modules). We discuss in detail the Kac
determinant formulae for the $S$- and $D$-modules
and the Kac and new determinant formulae for the $R$-module.
We also give a few low lying levels to clarify notations.

{}In section III we derive
the Kac and new determinant formulae using a generalization
\cite{ALT,CLT} of the BRST operator of Felder \cite{Felder}.
We also derive the relation between the Kac and new determinants. The
zeros of the Kac determinants were given in Ref \cite{open,ALT,CLT}. To
derive the Kac and new determinant formulae, we need to fix the orders of
these zeros, as well as the zero mode contribution. For level
$K\geq3$ FSCAs, the orders of some zeros of the Kac
and new determinants for a given module come from the string
functions that give the counting of states in a different module. This
is never the case for the Virasoro and superconformal algebras.

{}Next, we turn to the applications of the Kac determinant
formulae. In section IV, using our results for general $K$,
we deduce the values of the central charge $c$ and
conformal dimension $h$ for which FSCAs can have unitary
representations. In particular, we sketch the non-unitarity proof for the
$SU(2)$ minimal series and argue that the only unitary models are
precisely the known $SU(2)_K\otimes SU(2)_L/SU(2)_{K+L}$ coset models
\cite{GKO}.

{}Recent evidence supports the existence of the
so-called fractional superstrings (FSS) \cite{AT1}.
A particular representation (with $c=5$) of the
spin-4/3 FSS has bosons and fermions living in
three-dimensional flat Minkowski space-time \cite{AT2}.
In section V we prove
the no-ghost theorem for the space-time bosonic sector of this theory;
using the Kac determinant formulae for the $K=4$ FSCA,
we generalize Brower and Thorn's old
no-ghost theorem \cite{Thorn} and argue that the physical state
conditions remove all negative-norm states.

{}We also use
the Kac determinant formulae for the $K=4$ FSCA to examine the physical
null state structure of the critical spin-4/3 FSS
($c=10$) and find extra sets of zero-norm physical states;
this property is expected of any consistent string theory.

%============================================================================
\widetext
\newpage
\section{Kac and New Determinants}
\bigskip

{}In this section we present the Kac and new determinant formulae for an
arbitrary level $K$ fractional superconformal algebra (FSCA). We recover
the well-known Kac determinants for the Virasoro ($K=1$) and superconformal
($K=2$) algebras, and elaborate on
the case of the spin-4/3 parafermion current algebra ({\em i.e.}, the
$K=4$ FSCA) of Zamolodchikov and Fateev \cite{FZft}. We derive the Kac and
new determinant formulae in the next section.

\subsection{Preliminaries}
\medskip

{}By definition the Kac determinants for a given
algebra are representation independent.
The simplest representation of the level $K$ FSCA (convenient for
derivation of the Kac and new determinants) is
a non-interacting system of the ${\bf Z}_K$ parafermion
(PF) \cite{ZFpara} and a free boson.
The $SU(2)_K/U(1)$ coset model is a realization of
the ${\bf Z}_K$ PF theory. The chiral
$SU(2)_K$ WZW theory \cite{KZ} has central charge
\begin{equation}
 c_0={{3K}\over{K+2}}
\end{equation}
and consists of holomorphic Virasoro primary fields $\Phi^j_m(z)$
of conformal dimensions ${j(j+1)}\over{(K+2)}$.
The indices $j,m\in{\bf Z}/2$ label $SU(2)$ representations, where
\begin{equation}\label{restrictJM}
 0\leq j\leq K/2{}~{}~{\rm and}{}~{}~|m|\leq j{}~{}~{\rm with}{}~{}~j-m\in
{\bf Z}{}~.
\end{equation}
When we factor a $U(1)$ subgroup out of $SU(2)_K$, we correspondingly
factor the primary fields as
\begin{equation}\label{primfac}
 \Phi^j_m(z){}~={}~\phi^j_m(z)\,\exp\left\{i{m\over{\sqrt{K}}}
\,\varphi(z)\right\}{}~.
\end{equation}
Here $\varphi$ is the free $U(1)$ boson normalized so that
$\langle\varphi(z)\varphi(w)\rangle=-2{}~{\rm ln}(z-w)$.
$\phi^j_m(z)$ are the Virasoro primary
fields in the ${\bf Z}_K$ PF theory with conformal dimensions
\begin{equation}\label{pfdims}
 \Delta^j_m{}~={}~{j(j+1)\over (K+2)}{}~-{}~
{m^2\over K}{}~{}~{\rm for}{}~{}~|m|\leq j{}~.
\end{equation}

{}The $SU(2)_K$ algebra has an automorphism
\begin{equation}\label{automorphism}
\phi^j_m{}~={}~\phi^{K/2-j}_{m+K/2}{}~.
\end{equation}
With these identifications, we can consistently extend $\phi^j_m$ to $|m|>j$.

{}The fusion rules of the PF fields follow from those of the
$SU(2)_K$ theory:
\begin{equation}\label{fusionrules}
 [\phi^{j_1}_{m_1}]\otimes[\phi^{j_2}_{m_2}]{}~\sim
 \sum_{j=|j_1-j_2|}^r[\phi^j_{m_1+m_2}]{}~,
\end{equation}
where $r={\rm min}\{j_1+j_2{}~,{}~K-j_1-j_2\}$.

{}The $SU(2)_K$ currents are given by
\begin{eqnarray}\label{Jcurrents}
 J^\pm&=&{\sqrt{K}}\phi^0_{\pm1}\cdot\exp(\pm i\varphi/\sqrt{K}){}~,\\
 J^0&=&{i\over2}{\sqrt{K}}\partial\varphi{}~,
\end{eqnarray}
where $\phi^0_{\pm1}$ have conformal dimension $(K-1)/K$.

{}The $SU(2)_K$ primary fields read
\begin{equation}\label{SU(2)primary}
 \Phi^j_j (z)= \phi^j_j (z) \cdot\exp(ij\varphi (z)/\sqrt{K}){}~.
\end{equation}
Note $\Phi^j_{-j}=(\Phi^j_j)^\dagger$ are also primary.

{}Next we consider the so-called fractional supercurrent (whose conformal
dimension is $\Delta=(K+4)/(K+2)$)
\begin{equation}\label{WZWGcurrent}
 G=J^a_{-1}\Phi^1_a=
 {\sqrt{K}\over2}\epsilon\partial\varphi
 +{{iK}\over{K+4}}\eta{}~.
\end{equation}
Here $J^a_{-1}$ are the conformal dimension 1 creation modes of the $J^a (z)$
currents; $\epsilon\equiv\phi^1_0$ is the so-called
PF energy operator and $\eta$ is its
first PF descendent. The closed algebra generated
by the currents $T(z)$ and $G(z)$ is the level $K$ FSCA (Since
$\epsilon$ does not exist for $K=1$, $G(z)$ is absent in this case
and we recover the Virasoro algebra). The central charge of this
algebra is $c_0$.
If we turn on the background charge of the $\varphi$ boson,
the algebra, generated by the appropriately modified $T(z)$ and $G(z)$
currents, remains closed. We note that there exist other FSCAs
with additional fractional supercurrents \cite{AGT}.

{}The following OPEs define the level $K$ FSCA with arbitrary
central charge $c$:
\begin{eqnarray}\label{FSCA}
 &&T(z)T(w)={(c/2)\over(z-w)^4}+{2T(w)\over(z-w)^2}+
   {\partial T(w)\over(z-w)}+...{}~,\nonumber\\
 &&T(z)G(w)={\Delta\over(z-w)^2}G(w)+
   {1\over(z-w)}\partial G(w)+...{}~,\\
 &&G(z)G(w)= {(z-w)^{-2\Delta}}\left\{{c\over\Delta}+
   2{(z-w)^2}T(w)+...\right\}+\nonumber\\
 &&\qquad\lambda(c){(z-w)^{-\Delta}}\left\{G(w)
 +{1\over2}(z-w)\partial G(w)+...\right\}.\nonumber
\end{eqnarray}
The associativity condition for this algebra fixes the structure
constant $\lambda^2 (c)$:
\begin{equation}\label{lambda(c)}
 \lambda^2 (c)={2{K^2}{c_{111}^2}\over{3(K+4)^2}}(c_1-c){}~.
\end{equation}
Here $c_{111}$ is the $SU(2)_K$ structure constant for the OPE of two
chiral spin-1 primary fields to give a chiral spin-1 field and \cite{AGT}
\begin{equation}\label{c_1}
 c_1\equiv24/K+c_0{}~.
\end{equation}

{}As mentioned earlier, for $K=1$ the $G$ current is absent. For $K=2$,
$c_{111}=0$, $\lambda(c)=0$ and $G$ is the usual supercurrent. It is a
hermitian current of conformal dimension 3/2. For $K\geq3$,
$c_{111}$ is non-vanishing and
\begin{equation}\label{Gdagger}
G^\dagger =\alpha G{}~,
\end{equation}
where
\begin{equation}
 \alpha=\left\{
 \begin{array}{ll}
 +1 & {}~{}~{}~\mbox{if $c\leq c_1$}{}~,\\
 -1 & {}~{}~{}~\mbox{otherwise}{}~.
 \end{array} \right.
\end{equation}
For definiteness we will take the branch of Eqn.(\ref{lambda(c)}) that is
positive for $c<c_1$.

{}The vertex operators ${\chi^j_m}(0)$, labeled by the $SU(2)_K$ quantum
numbers $j$ and $m$, create states in modules of the level $K$ FSCA:
$\vert\chi^j_m \rangle\equiv\chi^j_m (0)\vert0\rangle$. The fields primary
with respect to $G(z)$ and $T(z)$
have the form $\chi^j_j (z)$ ($\chi^j_{-j} (z)=(\chi^j_j )^\dagger$ are
also primary).
$G(z)$ has $j=1$ and $m=0$ and
in terms of modings acts on
$\vert\chi^j_m \rangle$ in the same manner as
$\epsilon$ acts on the PF fields $\phi^j_m$:
\begin{eqnarray}\label{Gmoding}
 &&G(z){\chi_m^j}(0)={\sum_n}[z^{n+2j/(K+2)-1}G_{-n-2(j+1)/(K+2)}+
   z^{n-2/(K+2)-1}G_{-n}+\nonumber\\
 &&\qquad\qquad\qquad z^{n-2(j+1)/(K+2)}G_{-n+2j/(K+2)-1}]{\chi_m^j}(0){}~,
\end{eqnarray}
where the resulting states have
spins $j+1$, $j$ and $j-1$, respectively (The terms
not allowed by the fusion rules (\ref{fusionrules}) are
absent). The positive modes of the $T$ and $G$ currents annihilate the
highest weight states created by the primary fields:
\begin{equation}
 L_0 \vert h\rangle=h \vert h\rangle{}~,{}~{}~{}~L_n \vert h\rangle=
 G_r \vert h\rangle=0{}~,{}~{}~{}~n,r>0{}~.
\end{equation}
Here $r$ is in general a rational number.

{}The module $[\chi^j_j ]$ built from the highest weight state $\vert\chi^j_j
\rangle$ is spanned by the states
\begin{equation}\label{spanVerma}
 L^{\lambda_1}_{-n_1}...L^{\lambda_p}_{-n_p}
 G^{\rho_1}_{-r_1}...G^{\rho_q}_{-r_q}\vert\chi^j_j \rangle{}~,
\end{equation}
where the ordering is the same as for the states
\begin{equation}\label{spanSU(2)}
 a^{\lambda_1}_{-n_1}...a^{\lambda_p}_{-n_p}
 \epsilon^{\rho_1}_{-r_1}...\epsilon^{\rho_q}_{-r_q}\phi^j_j
 (0)\vert0\rangle{}~,{}~{}~{}~n_i>0{}~,{}~r_k\geq0{}~.
\end{equation}
Here $a_{-n_i}$ represent creation modes of the $\varphi$ boson, while
$\epsilon_{-r_k}$ stand for modes of the ${\bf Z}_K$ PF energy operator.
Note that $\lambda_i$ can be any non-negative integers, while $\rho_k$
can take only two values: zero and unity.

{}The naive counting of states
$\vert \chi^j_m \rangle$ of the form (\ref{spanVerma}) is identical to
the counting of states (\ref{spanSU(2)}) with the same quantum numbers
$j$ and $m$ and is given by the level $K$ string functions $C_{2m}^{2j}$
\cite{KP}
\begin{eqnarray}\label{stringfunctionsK}
 &&C^{2j}_{2m}={q^{d^j_m}\over{\eta^3}} {\sum_{r,s=0}^\infty} (-1)^{r+s}{q^
 {s(s+1)/2+r(r+1)/2+sr(K+1)}}\nonumber\\
 &&\times\left(q^{s(j-m)+r(j+m)}- q^{K+1-2j+s[K+1-j+m]+r[K+1-j-m]}\right).
\end{eqnarray}
Here $\eta$ is the Dedekind $\eta$-function
\begin{equation}\label{dede}
 \eta(q)=q^{1/24}\prod_{n=1}^\infty(1-q^n){}~,
\end{equation}
and
\begin{equation}
d^j_m=\Delta^j_m+{1\over{4(K+2)}}{}~.
\end{equation}

{}For later convenience we define the level of a state (\ref{spanVerma}) as
\begin{equation}\label{N}
N=N_0+\Delta^j_j{}~,
\end{equation}
where $\Delta^j_j$ is given by (\ref{pfdims}) and
\begin{equation}\label{N_0}
 N_0=\sum_{i=1}^p{}~\lambda_i n_i+\sum_{k=1}^q{}~ \rho_k r_k{}~
\end{equation}
is (generically a rational number) the level of the state
above the highest weight state $\vert\chi^j_j \rangle$.
The number of its descendents
({\em i.e.}, the number of states in the module
$[\chi^j_j ]$) at level $N$ is given
by $P_j (N)$, the latter being a coefficient in the power expansion
\begin{equation}\label{P}
 C_{2m} (q)\equiv\sum_{j} C^{2j}_{2m}
(q)=q^{-{{c_0}\over24}}\sum P_m (N)q^N{}~.
\end{equation}
The sum over $j$ includes its values allowed by the fusion
rules (\ref{fusionrules}) up to identifications (\ref{automorphism}),
{\em i.e.}, each independent string function appears once and only once.

{}We note that the $G$ current modings when acting on the highest
weight states $\vert\chi^j_j \rangle$ and $\vert\chi^{K/2-j}_{K/2-j}
\rangle$ ($0<j<K/4$) are the same. Hence the modules $[\chi^j_j]$ and
$[\chi^{K/2-j}_{K/2-j}]$ are structurally identical.

{}We end this subsection with the following remark. An important feature
of level $K\geq3$ FSCAs is the appearance
of cuts in the $GG$ OPE (\ref{FSCA}). Since
there are two different cuts on the right hand side, upon continuation of
a correlation function involving $G(z)G(w)$ OPE along a contour interchanging
$z$ and $w$ in the complex plane, it is inconsistent for the
correlator to pick up a simple
phase; that is, the current $G$ is
non-abelianly braided. There is only one FSCA, namely, the spin-4/3
($K=4$) parafermion current algebra, where it is possible to split the
fractional supercurrent so that the split currents are abelianly braided.
This makes for important technical simplifications. For example, it
enables one to write down the generalized commutation relations between
the modes of the split currents and to explicitly calculate
the Kac and new determinants. For other FSCAs the generalized
commutation relations have not yet been written down because of
technical difficulties arising from their non-abelian braiding properties.
Hence, the spin-4/3 parafermion current algebra provides
an important check on the Kac and new determinant formulae presented in
the next subsection, and we consider it in detail in subsection C.

\subsection{Kac and New Determinant Formulae}
\medskip

{}As mentioned earlier, so far the Kac determinant formulae have been known
only for the Virasoro and superconformal algebras. These Kac determinants
are polynomials in the highest weight $h$. One expects the determinants
of inner products of states in modules of FCSAs also be polynomials
in $h$. This is the case for some modules, but not for all.

{}There are three different types of modules in the Fock space of the
level $K$ FSCA.

{}({\em i}) In the module $[\chi^0_0 ]$, built from the highest
weight state $\vert\chi^0_0\rangle$, the counting of states
is given by $C_0$ defined in (\ref{P}). The fractional supercurrent
zero mode $G_0$ does not act on
$\vert\chi^0_0\rangle$ and the determinant of inner products of states at
any level is a polynomial in $h$.

{}({\em ii}) For $0<j\not=K/4$ the highest weight state
$\vert\chi^j_j\rangle$ is an eigenstate of $G_0$. The
counting of states in the module $[\chi^j_j]$ given
by $C_{2j}$. The determinant of inner products
of states at any level is a
polynomial in the $G_0$ eigenvalue, but is {\em not} a
polynomial in $h$. However, as mentioned earlier,
the modules $[\chi^j_j]$ and $[\chi^{K/2-j}_{K/2-j}]$ ($0<j<K/4$) are
structurally identical. This allows us to construct the highest weight state
$\vert h\rangle$ (as a linear combination of
$\vert\chi^j_j\rangle$ and $\vert\chi^{K/2-j}_{K/2-j}\rangle$) that is no
longer an eigenstate of $G_0$. The determinant of inner
products of states in the module
$[\chi^j_j \oplus\chi^{K/2-j}_{K/2-j}]$ built from $\vert h\rangle$
is a polynomial in $h$.
The number of states is given by $C_{2j} + C_{K-2j}(\equiv 2C_{2j})$.

{}({\em iii}) If $K$ is even, there is the third type of modules,
namely, the $[\chi^{K/4}_{K/4}]$ module. $G_0$
is not necessarily diagonal with respect to the highest weight state
$\vert\chi^{K/4}_{K/4}\rangle$. However, the determinant of
inner products of states is always
a polynomial in $h$. The counting of states is given by
$C_{K/2}$.

{}Now we define $\det({\cal M}^{(j)} (N))$ as the determinant of
inner products of states (\ref{spanVerma}) at level $N$.
It is a generalization of the Kac determinants for the Virasoro
and superconformal algebras. For the modules $[\chi^0_0 ]$,
$[\chi^{K/4}_{K/4}]$ and $[\chi^j_j \oplus\chi^{K/2-j}_{K/2-j}]$ ($0<j<K/4$)
the determinant is a polynomial in $h$, and we will refer
to it as the Kac determinant. For the module $[\chi^j_j]$ ($0<j\not=K/4$)
the determinant is a polynomial in the $G_0$ eigenvalue,
but is {\em not} a polynomial in $h$. We will refer
to these determinants as the new determinants. The Kac determinant
for the module $[\chi^j_j \oplus\chi^{K/2-j}_{K/2-j}]$ factorizes
into a product of the new determinants for the $[\chi^j_j]$ and
$[\chi^{K/2-j}_{K/2-j}]$ modules.

{}First we present (up to positive normalization constants independent of the
highest weight $h$ and the central charge $c$) the Kac and new determinant
formulae for the modules described above. Then we give explicit formulae for
the zeros of the Kac and new determinants. Finally, we comment on some
features of the determinants for FSCAs.

{}$\bullet${}~({\em i}) The module $[\chi^0_0]$.

{}The Kac determinant reads
\begin{equation}\label{kacdet1}
 \det({\cal M}^{(0)} ( N_0 ))=\alpha^{Q(0)}
 \prod_{r,s}[\alpha(h-h_{r,s})]^{P_\ell (N_0 -rs/K)}{}~,
\end{equation}
where $r,s\in{\bf N}$, $s-r=0{}~{\rm
mod}{}~K$ and $rs/K\leq N_0$; $2\ell=(s+r){}~{\rm mod}{}~K$ and
\begin{equation}
 Q(0) =\sum_n P_0 (N_0 -n^2){}~,
\end{equation}
where $n\in{\bf N}$, $n^2\leq {N_0}$.

{}$\bullet${}~({\em ii}) The module $[\chi^{K/4}_{K/4}]$, $K$ is even.

{}({\em a}) $G_0$ is diagonal with respect to the highest weight state
$\vert h;g^\pm \rangle$:
\begin{equation}
 G_0\vert h;g^\pm \rangle=g^\pm \vert h;g^\pm \rangle{}~.
\end{equation}
With the appropriate normalization of $G_0$, the eigenvalues $g^\pm$ are
given by
\begin{equation}\label{eigenK/4}
 g^\pm =\pm\sqrt{h-c/24}{}~.
\end{equation}

{}The Kac determinants for the modules built from $\vert h;g^\pm \rangle$ read
\begin{equation}\label{kacdet30}
 \det({\cal M}^{(K/4)}_{(\pm)} (N))=\alpha^{Q (K/4)}
 \prod_{r,s}(h-h_{r,s})^{{P_\ell} (N-rs/K)}{}~,
\end{equation}
where $N\equiv N_0+K/8(K+2)$, $s-r=K/2{}~{\rm mod}{}~K$ and $rs/K\leq N$;
$2\ell=(s+r){}~{\rm mod}{}~K$ and
\begin{equation}
 Q (K/4)=P_{K/4} (N)/2+\sum_n P_{K/4} (N-n^2){}~.
\end{equation}
Here $n^2\leq N$.

{}({\em b}) $G_0$ is {\em not} diagonal with respect to the highest
weight state $\vert h\rangle$.

{}The Kac determinant reads
\begin{equation}\label{kacdet3}
 \det({\cal M}^{(K/4)} (N))=\alpha^{Q(K/4)} (h-c/24)^{P_{K/4}(N)/2}
 \prod_{r,s}(h-h_{r,s})^{P_\ell (N-rs/K)}{}~,
\end{equation}
where $r,s,\ell$ and $N$ are the same as in Eqn.(\ref{kacdet30}).

{}$\bullet${}~({\em iii}) The module $[\chi^j_j]$, $0<j\not=K/4$.

{}({\em a}) The highest weight state
$\vert\chi^j_j\rangle\equiv\vert h;g^j\rangle$ is an eigenstate of $G_0$:
\begin{equation}
 G_0\vert h;g^j \rangle=g^j \vert h;g^j \rangle{}~.
\end{equation}
With the appropriate normalization of $G_0$, the eigenvalue $g^j$ is given by
\begin{equation}\label{g^pm}
 g^j =a^j \sqrt{c_1-c}+{\rm sgn}(j-K/4)\sqrt{h-h^j_0}{}~,
\end{equation}
where
\begin{equation}
 a^j={|K-4j|\over{\sqrt{24}(K+4)}}{}~,
\end{equation}
and we define for arbitrary $j$
\begin{equation}
 h^j_0\equiv{{c-c_0}\over24}+\Delta^j_j=
 {c\over24}-{{(K-4j)^2}\over{8K(K+2)}}{}~.
\end{equation}

{}The new determinant reads
\begin{equation}\label{kacdet-g}
\det({\cal M}^{(j)} (N))=\alpha^{Q(j)}
 \prod_{r,s} [\alpha(g^j -g_{r,s})(g^j -g_{s,r})]^{P_\ell (N-rs/K)}{}~,
\end{equation}
where $N=N_0+\Delta^j_j$, $s-r=2j{}~{\rm mod}{}~K$ and $rs/K\leq
N$; $2\ell=(s+r){}~{\rm mod}{}~K$ and
\begin{equation}
 Q(j)=\sum_n P_j (N-n^2){}~.
\end{equation}

{}({\em b}) The module $[\chi^j_j \oplus\chi^{K/2-j}_{K/2-j}]$,
$0<j<K/4$.

{}The highest weight state $\vert h\rangle$ is a linear combination of
$\vert\chi^j_j\rangle$ and $\vert\chi^{K/2-j}_{K/2-j}\rangle$ such that
$G_0$ is {\em not} diagonal with respect to $\vert h\rangle$.

{}The Kac determinant reads
\begin{equation}\label{kacdet2}
 \det({\cal M}^{(j)}_0 (N))=[\alpha
 (h- h^j_0)]^{P_j (N)}
 \prod_{r,s}[(h-h_{r,s})(h-h_{s,r})]^{P_\ell (N-rs/K)}{}~,
\end{equation}
where $r,s,\ell$ are the same as in (\ref{kacdet-g}).

{}The Kac determinant (\ref{kacdet2}) factorizes as follows:
\begin{equation}\label{hgdet}
 \det({\cal M}^{(j)}_0 (N))=[\alpha
 (h-h^j_0 )]^{P_j (N)}
 \det({\cal M}^{(j)} (N))
 \det({\cal M}^{(K/2-j)} (N)){}~.
\end{equation}

{}The zeros of the Kac determinants (\ref{kacdet1}), (\ref{kacdet30}),
(\ref{kacdet3}) and (\ref{kacdet2}) read \cite{open}
\begin{equation}\label{nullstates}
 h_{r,s}={{c-c_0}\over24}+\Delta^j _j +{1\over96}\left((r+s)\sqrt{c_0-c}
 +(r-s)\sqrt{c_1-c}\right)^2=h^j_0+b^2_{r,s}{}~,
\end{equation}
where
\begin{equation}
 b_{r,s}={1\over{\sqrt{96}}}\left((r+s)\sqrt{c_0-c}+
 (r-s)\sqrt{c_1-c}\right){}~.
\end{equation}
The zeros of the new determinants (\ref{kacdet-g}) are given by
\begin{equation}\label{g_rs1}
 g_{r,s}=a^j \sqrt{c_1-c}+{\rm sgn}(j-K/4)b_{r,s}{}~.
\end{equation}
Here $2j=(s-r){}~{\rm mod}{}~K$. The relation
between the zeros of the Kac and corresponding new determinants reads
\begin{equation}
 (g^j -g_{r,s})(g^{K/2-j} -g_{r,s})=-(h-h_{r,s}){}~.
\end{equation}
Their orders $P_m (N)$ are given by
\begin{equation}
 C_{2m} (q)\equiv\sum_{j} C^{2j}_{2m} (q)=q^{-{K\over{8(K+2)}}}\sum
P_m (N)q^N{}~.
\end{equation}
The sum over $j$ includes its values allowed by the fusion
rules (\ref{fusionrules}) up to identifications (\ref{automorphism}),
{\em i.e.}, each independent string function appears once and only once.

{}We conclude this subsection with the following remarks:

{}({\em a}) We define the Kac
and new determinants for a linearly independent set of states
where the total number of the $T$ and $G$ modes is minimal. If this number
is not minimal, the determinant will be different
by an overall factor (that can be either positive or
negative) independent of $h$, but in general dependent on $c$.

{}({\em b}) For the Virasoro algebra the $G$ current is absent, while for
the superconformal algebra it is always hermitian. Therefore, for the $K=1,2$
cases $\alpha\equiv1$ in all of the above formulae. For
level $K\geq3$ FSCAs, $\alpha$ flips sign as the central charge exceeds
$c_1$ defined in (\ref{c_1}). Hence in
the Kac and new determinants there appears the factor of $\alpha$ to the power
(defined mod 2) determined by the total number of the $G$ current modes
in a linearly independent set of states with minimal counting.

{}({\em c}) For level $K\geq3$ FSCAs, the orders of some zeros of the Kac
and new determinants for a given module come from the string
functions that give the counting of states in a different module. This
is never the case for the Virasoro and superconformal algebras.

{}The examples in the next subsection illuminate these issues.

\subsection{Examples}
\medskip

{}Now we consider some concrete cases in order to clarify our conventions
and notations. We give the Kac and new
determinant formulae up to normalization constants
independent of the highest weight $h$ and the central charge $c$.

{}$\bullet${}~({\em i}) The Virasoro algebra ($K=1$).

{}We have only one module
$[\chi^0_0]$ and the counting of states is given by the string function
\begin{equation}
 C_0 \equiv C^0_0=1/\eta(q)=q^{-1/24} \prod_{n=1}^{\infty} (1-q^n )^{-1} {}~.
\end{equation}
Eqn.(\ref{kacdet1}) (with $\alpha\equiv1$) reduces to the well-known
Kac determinant formula for the Virasoro algebra \cite{Kac}.

{}$\bullet${}~({\em ii}) The superconformal algebra ($K=2$).

{}({\em a}) The Neveu-Schwarz module, or the $[\chi^0_0]$ module. According
to Eqn.(\ref{P}) the counting of states is given by
\begin{equation}
 C_0 \equiv C^0_0 + C^2_0 = q^{-1/16} \prod_{n=1}^\infty
 {{1+q^{n-1/2}}\over{1-q^n}}{}~,
\end{equation}
Eqn.(\ref{kacdet1}) (again with $\alpha\equiv1$) now reduces to
the Kac determinant formula for the Neveu-Schwarz module.

{}({\em b}) The Ramond module, or the $[\chi^{1/2}_{1/2}]$ module. This
is the simplest $[\chi^{K/4}_{K/4}]$ module. The counting is given
by the string function
\begin{equation}
 C_1 \equiv C^1_1=q^{-1/16} \left( q^{1/16} \prod_{n=1}^\infty
 {{1+q^n}\over{1-q^n}} \right){}~,
\end{equation}
The level of a state in this module according to Eqn.(\ref{N}) is defined
as $N=N_0 +1/16$, where $N_0$ is the level of the state above the highest
weight state. Eqn.(\ref{kacdet3}) is then the Kac determinant
formula for the Ramond module \cite{FQS2}.

{}To illuminate the Kac determinant formula (\ref{kacdet30}) for the
$[\chi^{K/4}_{K/4}]$ module, we consider the Ramond module built
from an eigenstate $\vert h;g^\pm\rangle$ of the supercurrent zero mode $G_0$.
Due to the zero mode relation
\begin{equation}\label{0SCA}
 G^2_0=L_0-c/24{}~,
\end{equation}
we have
\begin{equation}
 g^\pm =\pm\sqrt{h-c/24}{}~.
\end{equation}

{}A linearly independent set of states at level $17/16=1+1/16$ reads
\begin{equation}
 \vert\phi^\pm_1\rangle=L_{-1} \vert h;g^\pm\rangle{}~,{}~{}~{}~
 \vert\phi^\pm_2\rangle=G_{-1}\vert h;g^\pm\rangle{}~.
\end{equation}
The Kac determinants of these states can be computed using the superconformal
mode algebra:
\begin{equation}
 \det({\cal M}_{(\pm)} )=4(h-h_{1,2})(h-h_{2,1}){}~,
\end{equation}
where $h_{1,2}$ and $h_{2,1}$ are given by Eqn.(\ref{nullstates})
at $K=2$ and $j=1/2$. Note that at this level the Kac
determinant is a polynomial in $h$. This holds at all levels for all
the $[\chi^{K/4}_{K/4}]$ modules.

{}$\bullet${}~({\em iii}) The spin-4/3 parafermion current algebra ($K=4$).

{}The chiral fractional supercurrent
$G(z)$, whose conformal dimension is $4/3$, can be split into two pieces:
\begin{equation}
 G(z)=\left\{
 \begin{array}{ll}
 G^+ (z)+G^- (z) & {}~{}~{}~\mbox{if $c\leq 8$}{}~,\\
 i(G^- (z)-G^+ (z)) & {}~{}~{}~\mbox{otherwise}{}~.
 \end{array} \right.
\end{equation}
The currents $G^\pm (z)$ are abelianly braided ({\em i.e.}, parafermionic),
and with the $T(z)$ current form the closed spin-$4/3$
parafermion current algebra of Zamolodchikov and Fateev \cite{FZft}
\begin{eqnarray}\label{OPE}
  G^{\pm}(z)G^{\pm}(w)&=&{\lambda^{\pm}\over(z-w)^{4/3}}\left\{{G^{\mp}(w)
   +{1\over2}(z-w)\partial G^{\mp}(w)}+...\right\},\nonumber\\
  G^{+}(z)G^{-}(w)&=&{1\over(z-w)^{8/3}}\left\{{3c\over8}+
   {{(z-w)^2}T(w)}+...\right\},\\
  T(z)G^\pm(w)&=&{(4/3)G^\pm(w)\over(z-w)^2}+
   {\partial G^\pm(w)\over(z-w)}+...{}~,\nonumber\\
  T(z)T(w)&=&{(c/2)\over(z-w)^4}+{2T(w)\over(z-w)^2}+
   {\partial T(w)\over(z-w)}+...{}~,\nonumber
\end{eqnarray}
where
\begin{equation}
 \lambda^+ =\alpha\lambda^-=\lambda={\sqrt{|8-c|}\over\sqrt{6}}{}~,{}~{}~{}~
 \alpha=\left\{
 \begin{array}{ll}
 +1 & {}~{}~{}~\mbox{if $c\leq 8$}{}~,\\
 -1 & {}~{}~{}~\mbox{otherwise}{}~.
 \end{array} \right.
\end{equation}

{}The $G^\pm$ currents have the following hermiticity properties:
\begin{equation}\label{Gdagger1}
 (G^\pm)^\dagger =\alpha G^\mp{}~.
\end{equation}

{}The algebra (\ref{OPE}) obeys a ${\bf Z}_3$ symmetry. The
$G^{\pm}(z)$ and $T(z)$ currents have ${\bf Z}_3$ charges $q=\pm1$ and zero,
respectively. The ${\bf Z}_3$ charge $q$ is defined mod 3.

{}The modes of the currents
\begin{equation}
 T(z)=\sum_n z^{-n-2}L_n{}~,{}~{}~{}~G^\pm (z)=\sum_r z^{-r-4/3}G^\pm _r
\end{equation}
satisfy the commutation relations
\begin{eqnarray}\label{LGcommutators}
 \left[L_m,L_n\right]&=&(m-n)L_{m+n}
 +{c\over12}(m^3-m)\delta_{m+n}{}~,\nonumber\\
 \left[L_m,G^\pm_r\right]&=&
 \left({m\over3}-r\right)G^\pm_{m+r}{}~.
\end{eqnarray}

{}The modings of the currents $G^{\pm}(z)$ and the generalized commutation
relations (GCRs) for them depend on a representation of the algebra.
First we consider the representations that correspond to the
integer spin $j$ modules. The Fock space falls into sectors
${\cal H}_q$ labeled by their ${\bf Z}_3$ charge. The currents $G^\pm$
act on the Fock space sectors according to the rules
\begin{equation}\label{FockSpace}
  G^{\pm}: {\cal H}_q\rightarrow{\cal H}_{q\pm1}{}~.
\end{equation}
With these actions, the mode expansions of the $G^{\pm}$ currents are
defined as
\begin{equation}\label{Gmodes}
  G^{\pm}(z)\chi_q(0)=\sum_n z^{n\mp q/3}G^{\pm}_{-1-n-(1\mp q)/3}\chi_q(0){}~,
\end{equation}
where $\chi_q$ is an arbitrary state in ${\cal H}_q$.  These
mode expansions can be inverted to give
\begin{equation}\label{Gmodes1}
  G^{\pm}_{n-(1\mp q)/3}\chi_q(0)=\oint_\gamma{{\rm d}z
   \over2\pi i}{}~z^{n\pm q/3}G^{\pm}(z)\chi_q(0){}~.
\end{equation}
Here, $\gamma$ is a contour encircling the origin once,
where $\chi_q(0)$ is inserted.

{}The $G_r$ modes satisfy the following GCRs
\begin{eqnarray}
 &&\sum_{\ell=0}^\infty c^{(1/3)}_\ell\left[
  G^\pm_{\pm{q\over3}+n-\ell}G^\pm_{{2\pm q\over3}+m+\ell}+
  G^\pm_{\pm{q\over3}+m+1-\ell}G^\pm_{{2\pm q\over3}+n-1+\ell}\right]{}~=
 {\lambda^\pm} G^\mp_{{2\pm2q\over3}+n+m}{}~,\label{GCR1a}\\
 &&\sum_{\ell=0}^\infty c^{(-1/3)}_\ell\left[
  G^{+}_{{1+q\over3}+n-\ell}G^{-}_{-{1+q\over3}+m+\ell}+
 G^{-}_{-{2+q\over3}+m-\ell}G^{+}_{{2+q\over3}+n+\ell}\right]{}~=\nonumber\\
 &&\qquad\qquad\qquad L_{n+m}+{3c\over16} \left(n+1+{q\over3}\right)
  \left(n+{q\over3}\right)\delta_{n+m}{}~,\label{GCR1b}
\end{eqnarray}
when acting on a state with the ${\bf Z}_3$ charge $q$.
Here $c^{(\alpha)}_\ell$ are the binomial coefficients
\begin{equation}\label{binomialcoefficients}
 (1-x)^\alpha=\sum_{\ell=0}^{\infty}{c^{(\alpha)}_{\ell}}x^{\ell}{}~.
\end{equation}

{}According to the hermiticity assignments (\ref{Gdagger1})
and the mode expansions (\ref{Gmodes}), the hermiticity properties of
the $G^{\pm}_r$ modes are $(G^+_r)^\dagger=\alpha G^-_{-r}$.

{}The highest weight states $\vert{h;q}\rangle$ with the ${\bf Z}_3$
charge $q$ satisfy the following conditions
\begin{equation}
 L_0\vert{h;q}\rangle=h\vert{h;q}
\rangle{}~,{}~{}~{}~L_n\vert{h;q}\rangle=0{}~,{}~{}~{}~G^{\pm}
 _{n-(1\mp q)/3}\vert{h;q}\rangle=0{}~,{}~{}~{}~n>0{}~.
\end{equation}

{}Next we consider the representations of the algebra corresponding to
the half odd integer spin $j$ modules. The fields
$z^{4/3} G^{\pm}(z) {\chi_R}(0)$ (where $\chi_R$ is an arbitrary state in
the Fock space) are
double-valued analytic functions, {\em i.e.}, the fields
$\chi_R (z)$ create square root cuts in the complex plane.
For the purpose of
calculating the Kac and new determinants it suffices to consider
only the states ${\chi_R}(0)$ that are ${\bf Z}_3$ singlets.
The modings of the $G^{\pm}$ currents then are given by the following OPEs
\begin{equation}\label{GmodesR}
 G^{\pm}(z){\chi_R}(0)=2^{-{2\over3}}\sum_{n=-\infty}^{\infty}z^{n/2-4/3}
 G^{\pm}_{-n/2}{\chi_R}(0){}~,{}~{}~{}~n\in{\bf Z}{}~,
\end{equation}
where the overall normalization is chosen for later convenience.
These mode expansions can be inverted to give
\begin{equation}
 G^{\pm}_{n/2}{\chi_R}(0)=2^{-{1\over3}}{\oint_\gamma} {{\rm d}z\over2\pi
 i}{}~z^{1/3+n/2}G^{\pm}(z){\chi_R}(0){}~.
\end{equation}
Here, $\gamma$ is a contour encircling the origin twice, where
$\chi_R(0)$ is inserted.

{}The modes $G^{\pm}_{n/2}$ are related to each other via
$G^+_{n/2}=(-1)^n \alpha G^-_{n/2}$ due to the single-bypass relations
discussed in Ref \cite{FZft}.  Defining $\tilde G_{n/2}\equiv G^-_{n/2}$,
we find that the modes $\tilde G_{n/2}$ satisfy the following GCR
\begin{eqnarray}\label{GCR2}
 &&\sum_{\ell=0}^\infty D_{({1\over3},-{1\over3})}^{(\ell)}\left[
  \tilde G_{{n-\ell}\over2} \tilde G_{{m+\ell}\over2}+
  \tilde G_{{m-\ell}\over2} \tilde G_{{n+\ell}\over2}\right]{}~={}~
  {\lambda\over2} (-1)^{n+m} \tilde G_{{n+m}\over2}+\nonumber\\
 &&\qquad\alpha[(-1)^n+(-1)^m]\left\{L_{{n+m}\over2}+{3c\over8}
\left({n^2\over8}-{5\over48}\right)\delta_{n+m}\right\}{}~,
\end{eqnarray}
where $D_{(\alpha,\beta)}^{(\ell)}$ are the binomial coefficients
\begin{equation}
 (1-x)^\alpha (1+x)^\beta=\sum_{\ell=0}^{\infty}D_{(\alpha,\beta)}^{(\ell)}
 x^{\ell}{}~.
\end{equation}

{}According to the hermiticity assignments (\ref{Gdagger1}) and the mode
expansions (\ref{GmodesR}), the hermiticity properties of the $\tilde
G_{n/2}$ modes are $(\tilde G_{n/2})^\dagger=(-1)^n \tilde G_{-n/2}$.

{}The highest weight states satisfy the following conditions
\begin{equation}
 L_0\vert h\rangle_R =h\vert h\rangle_R{}~,{}~{}~{}~
L_n\vert h\rangle_R =0{}~,{}~{}~{}~
 \tilde G_{n/2}\vert h\rangle_R =0{}~,{}~{}~{}~n>0{}~.
\end{equation}

{}Now we turn to the description of the spin-4/3 parafermion
current algebra modules where the counting of states is given by
the $K=4$ string functions \cite{KP,Yang}
\begin{eqnarray}\label{stringfunctions}
 C^0_0+C^4_0
&=& {1\over{\eta^2}(q)}\left(\sum_{n=-\infty}^\infty
  q^{3n^2} \right){}~,\nonumber\\
 C^2_2
&=& {q^{1/12}\over{\eta^2}(q)}\left(\sum_{n=-\infty}^\infty q^{3n^2+n}
  \right){}~,\nonumber\\
 C^2_0
&=& {q^{1/3}\over{\eta^2}(q)}\left(\sum_{n=-\infty}^\infty q^{3n^2+2n}
  \right){}~,\\
 C^4_2=C^0_{2}
&=& {q^{3/4}\over{\eta^2}(q)}\left(\sum_{n=0}^\infty
  q^{3n^2+3n} \right){}~,\nonumber\\
 C^1_1+C^3_1
&=& C^3_3+C^1_3={1/{\eta(q^{1\over2})}}{}~.\nonumber
\end{eqnarray}

According to Eqn.(\ref{P}) we define
\begin{eqnarray}\label{counting4}
 &&C_0\equiv C^0_0+C^2_0+C^4_0=q^{-1/12}\sum P_0 (N)q^N=q^{-1/12}
 (1+q^{1/3}+...){}~,\nonumber\\
 &&C_2\equiv C^0_2+C^2_2+C^4_2=q^{-1/12}\sum P_1 (N)q^N=q^{-1/12}
 (q^{1/12}+2q^{3/4}+...){}~,\\
 &&C_1\equiv C^1_1+C^3_1=q^{-1/12}\sum P_{1/2} (N)q^N=q^{-1/12}
 (q^{1/16}+q^{9/16}+...){}~,\nonumber\\
 &&C_3\equiv C^1_3+C^3_3=q^{-1/12}\sum P_{3/2} (N)q^N=C_1{}~.\nonumber
\end{eqnarray}
We also define the following combination of the string functions:
\begin{equation}\label{combo}
 C_0+C_2=q^{-1/12}\sum P(N)q^N{}~.
\end{equation}

{}({\em a}) The S-module, or the $[\chi^0_0]$ module. This module
is built from the highest weight state
$\vert{h;0}\rangle$. We define the level of a state
as $N=N_0$, where $N_0$ is the level of the state above
$\vert{h;0}\rangle$. Eqn.(\ref{Gmodes}) gives $N\in{\bf Z}$ or
$N\in{\bf Z}+1/3$. The $S$-module falls into three
submodules, $S^{(0)}$ and $S^{(\pm)}$, labeled according to the ${\bf
Z}_3$ charge of the states. The $S^{(0)}$-submodule
contains the states (with quantum numbers $j=0,2$ and $ m=0$) at integer
levels, while the $S^{(\pm)}$-submodules
consist of the states ($j=1$ and $m=0$) at levels $N\in{\bf Z}+1/3$.
The number of states at level $N$ in each submodule is given by
$P_0 (N)$ defined in (\ref{counting4}).

{}For the $S$-module we define the Kac determinant at integer levels
as that of the $S^{(0)}$-submodule, and at levels $N\in{\bf Z}+1/3$
as that of the $S^{(-)}$-submodule. The Kac
determinants for the $S^{(+)}$- and $S^{(-)}$-submodules are identical.
We choose the normalization $\langle{h;0}\vert{h;0}\rangle=1$.

{}The Kac determinant formula for the $S$-module is given by
\begin{equation}\label{kacdet-S}
 \det({\cal M}^S_N)=
 \alpha^{Q(S)}\prod_{r,s}(h-h^{S}_{r,s})^{P
 (N-rs/4)}{}~,
\end{equation}
where $r,s\in{\bf N}$,
$s-r=0{}~{\rm mod}{}~4$, $rs/4\leq N$, and
\begin{equation}
 Q(S)=\sum_n P_0 (N-{n^2}/4){}~.
\end{equation}
Here $n\in{\bf N}$ and ${n^2}/4\leq N$.
The zeros $h^S_{r,s}$ of the Kac determinant (\ref{kacdet-S})
are given by Eqn.(\ref{nullstates})
\begin{equation}\label{h-S}
 h^S_{r,s}={{c-2}\over24}+{1\over96}\left((r+s)\sqrt{2-c}+(r-s)\sqrt{8-c}
\right)^2{}~.
\end{equation}

{}Note that the order of the zeros $P(N-rs/4)$ comes from the sum
$C_0+C_2$ defined in Eqn.(\ref{combo}), not just from $C_0$.

{}We present a few low lying levels explicitly to clarify
the notation. There is $P_0 (1/3)=1$ state in the $S^{(-)}$-submodule at
level 1/3:
\begin{equation}\label{S_1/3}
 \vert\phi\rangle=G^-_{-{1\over3}}\vert{h;0}\rangle{}~,
\end{equation}
All the other states at this level are linearly dependent
on (\ref{S_1/3}). For instance, using the GCR (\ref{GCR1a}) we have
${G^+_0}G^+_{-{1\over3}}\vert{h;0}\rangle=({\lambda^+}/2)
G^-_{-{1\over3}}\vert{h;0}\rangle$, and we choose the
state (\ref{S_1/3}), since it has the minimal number of the
$G$ creation operators. The Kac determinant can be calculated
using (\ref{LGcommutators}) and (\ref{GCR1b}):
\begin{equation}
 \det({\cal M}^S_{1/3})=\alpha(h-h_{1,1}){}~.
\end{equation}
At level 1 there are $P_0 (1)=2$ states in the $S^{(0)}$-submodule
\begin{equation}
 \vert\phi_1\rangle=L_{-1}\vert{h;0}\rangle{}~,{}~{}~{}~\vert\phi_2\rangle=G^-_
 {-{2\over3}}G^+_{-{1\over3}}\vert{h;0}\rangle{}~.
\end{equation}
The Kac determinant can be obtained using the commutation relations
(\ref{LGcommutators}) and GCRs (\ref{GCR1a}) and (\ref{GCR1b}):
\begin{equation}
 \det({\cal M}^S_1)={4\over3}(h-h_{1,1})^2(h-h_{2,2}){}~.
\end{equation}
At the next level $N=4/3$ there are $P_0 (4/3)=3$ states
\begin{equation}
 \vert\phi_1\rangle=L_{-1}G^-_{-{1\over3}}\vert{h;0}\rangle{}~
 ,{}~\vert\phi_2\rangle=G^+_{-1}G^+_{-{1\over3}}\vert{h;0}\rangle{}~,{}~
 \vert\phi_3\rangle=G^-_{-{4\over3}}\vert{h;0}\rangle{}~,
\end{equation}
and the Kac determinant is given by
\begin{equation}
 \det({\cal M}^S_{4/3})={4\over3}
 (h-h_{1,1})^2(h-h_{2,2})(h-h_{1,5})
 (h-h_{5,1}){}~.
\end{equation}

{}({\em b}) The $D$-module, or the $[\chi^1_1]$ module.
This module (of the $[\chi^{K/4}_{K/4} ]$ type) is built from
the highest weight state $\vert{h;-1}\rangle$.
We define the level of a state as $N=N_0+1/12$, where
$N_0$ is the level of the state above
$\vert{h;-1}\rangle$. Eqn.(\ref{Gmodes}) gives
$N\in{\bf Z}+1/12$ or $N\in{\bf Z}+3/4$.
The $D$-module falls into three submodules,
$D^{(0)}$ and $D^{(\pm)}$. The $D^{(\pm)}$-submodules consist of the
states ($j=1$ and $m=1$)
at levels $N\in{\bf Z}+1/12$, while the $D^{(0)}$-submodule contains the
states ($j=0,2$ and $m=1$) at levels $N\in{\bf Z}+3/4$. The number of
states at level $N$ in
each submodule is given by $P_1 (N)$ defined in (\ref{counting4}).

{}For the $D$-module we define the Kac determinant at levels $N\in{\bf
Z}+1/12$ as that of the $D^{(-)}$-submodule, and at levels $N\in{\bf
Z}+3/4$ as that of the $D^{(0)}$-submodule. The Kac determinants for the
$D^{(+)}$- and $D^{(-)}$-submodules are identical for $N>1/12$.
We choose the normalization $\langle{h;-1}
\vert{h;-1}\rangle=1$, $G^-_0\vert{h;-1}\rangle=\sqrt{h-c/24}\vert{h;+1}
\rangle$.

{}The Kac determinant formula for the $D$-module is given by
\begin{equation}\label{kacdet-D}
 \det({\cal M}^D_N)=
 \alpha^{Q(D)}(h-c/24)^{P_1 (N)/2}\prod_{r,s}
 (h-h^{D}_{r,s})^{P(N-{rs/4})}{}~.
\end{equation}
where $s-r=2{}~{\rm mod}{}~4$, $rs/4\leq N>1/12$ and
\begin{equation}
 Q(D)=P_1 (N)/2+\sum_n P_1 (N-n^2){}~.
\end{equation}
Here $n^2\leq N$.
The zeros $h^D_{r,s}$ are given by
\begin{equation}\label{h-D}
 h^D_{r,s}={c\over24}+{1\over96}\left((r+s)\sqrt{2-c}+(r-s)\sqrt{8-c}
\right)^2{}~.
\end{equation}

{}For example, at level $3/4=2/3+1/12$ there are $P_1 (3/4)=2$ states in
the $D^{(0)}$-submodule:
\begin{equation}
 \vert\phi_1\rangle=G^+_{-{2\over3}}\vert{h;-1}
\rangle{}~,{}~{}~{}~\vert\phi_2\rangle
 =G^-_{-{2\over3}}G^-_0\vert{h;-1}\rangle{}~,
\end{equation}
and
\begin{equation}
 \det({\cal M}^D_{3/4})=\alpha(h-c/24)(h-h_{1,3})(h-h_{3,1}){}~.
\end{equation}
At level $13/12=1+1/12$ there are $P_1 (13/12)=2$ states in the
$D^{(-)}$-submodule
\begin{equation}
 \vert\phi_1\rangle=L_{-1}\vert{h;-1}\rangle{}~,{}~{}~{}~
 \vert\phi_2\rangle=G^+_{-1}G^-_0\vert{h;-1}\rangle{}~.
\end{equation}
and the Kac determinant reads
\begin{equation}
 \det({\cal M}^D_{13/12})={4\over3}(h-c/24)(h-h_{1,3})(h-h_{3,1}){}~.
\end{equation}

{}We also give the first non-trivial level in the $D$-module built from
the highest weight states that are the eigenstates of the fractional
supercurrent zero mode $G_0$:
\begin{equation}
 G_0 \vert h;g^\pm \rangle = g^\pm \vert h;g^\pm \rangle{}~.
\end{equation}
With our normalization of $G_0$ in
Eqn.(\ref{eigenK/4}),
\begin{equation}
 G_0=\left\{
 \begin{array}{ll}
 G^+_0 +G^-_0 & {}~{}~{}~\mbox{if $c\leq 8$}{}~,\\
 i(G^-_0 -G^+_0 ) & {}~{}~{}~\mbox{otherwise}{}~.
 \end{array} \right.
\end{equation}
The eigenstates of $G_0$ read
\begin{equation}
 \vert h;g^\pm\rangle=\left\{
 \begin{array}{ll}
 {1\over\sqrt{2}}(\vert h;-1\rangle\pm\vert h;+1\rangle) &
 {}~{}~{}~\mbox{if $c\leq 8$}{}~,\\
 {1\over\sqrt{2}}(\vert h;-1\rangle\pm i\vert h;+1\rangle) &
 {}~{}~{}~\mbox{otherwise}{}~.
 \end{array} \right.
\end{equation}
They correspond to the eigenvalues $g^\pm =\pm\sqrt{h-c/24}$.
At level 3/4 we have the following states
\begin{equation}
 \vert\phi^\pm_1\rangle=G^+_{-{2\over3}}\vert
 h;g^\pm\rangle{}~,{}~{}~{}~\vert\phi^\pm_2\rangle=G^-_{-{2\over3}}G^-_0\vert
 h;g^\pm\rangle{}~.
\end{equation}
The corresponding Kac determinants read
\begin{equation}
 \det({\cal M}^{(\pm)}_{3/4})=\alpha(h-h_{1,3})(h-h_{3,1}){}~.
\end{equation}

{}({\em c}) The $R$-module, or the $[\chi^{1/2}_{1/2}
\oplus\chi^{3/2}_{3/2}]$ module. This module is built from the
highest weight state $\vert h\rangle_R$ that is {\em not} an eigenstate
of the $G_0$ operator.
We define the level of a state in this module as
$N=N_0+1/16$, where $N_0$ is the level of the state above  $\vert
h\rangle_R$. Eqn.(\ref{GmodesR}) gives $N\in{\bf
Z}/2+1/16$ (At integer levels $j=1/2$ and $m=1/2$, or $j=3/2$ and
$m=3/2$, while at half odd integer levels $j=3/2$ and $m=1/2$,
or $j=1/2$ and $m=3/2$). The number of states at level $N$ is given by
$P_{1/2} (N)+P_{3/2} (N)=2P_{1/2} (N)$ defined in (\ref{counting4}).
We choose the normalization ${_R}\langle h\vert h\rangle_R =1$.

{}The Kac determinant formula for the $R$-module is given by
\begin{equation}\label{kacdet-R}
 \det({\cal M}^R_N)=
 [\alpha(h+1/48-c/24)]^{P_{1/2} (N)}\prod_{r,s}
 [(h-h^{R}_{r,s})(h-h^{R}_{s,r})]^{P_{1/2} (N-{rs/4})}{}~,
\end{equation}
where $s-r=1{}~{\rm mod}{}~4$.
The zeros $h^R_{r,s}$ read
\begin{equation}\label{h-R}
 h^R_{r,s}={c\over24}-{1\over{48}}+{1\over96}\left((r+s)\sqrt{2-c}+(r-s)
 \sqrt{8-c}\right)^2{}~.
\end{equation}

{}We give a few low lying states in the $R$-module. At
level $1/16$ there are $2P_{1/2} (1/16)$=2 states
\begin{equation}
 \vert\phi_1\rangle=\vert h\rangle_R,{}~{}~{}~
\vert\phi_2\rangle=\tilde G_0\vert
 h\rangle_R{}~,
\end{equation}
and the Kac determinant can be calculated using Eqn.(\ref{LGcommutators})
and GCR (\ref{GCR2})
\begin{equation}
 \det({\cal M}^R_{1/16})=\alpha(h+1/48-c/24){}~.
\end{equation}
At level $9/16=1/2+1/16$ there are $2P_{1/2} (9/16)=2$ states
\begin{equation}
 \vert\phi_1\rangle=\tilde G_{-{1\over2}}\vert h\rangle_R{}~,
 {}~{}~{}~\vert\phi_2\rangle=
 \tilde G_{-{1\over2}} \tilde G_0\vert h\rangle_R{}~.
\end{equation}
The Kac determinant at this level is given by
\begin{equation}\label{kacdet-R-9/16}
 \det({\cal M}^R_{9/16})={16\over9}\alpha(h+1/48-c/24)
 (h-h_{1,2})(h-h_{2,1}){}~.
\end{equation}
At level $17/16=1+1/16$ we have $2P_{1/2} (17/16)=4$ states
\begin{eqnarray}
 \vert\phi_1\rangle&=&L_{-1}\vert h\rangle_R{}~,{}~{}~{}~\vert\phi_2\rangle=
 L_{-1}\tilde G_0\vert h\rangle_R{}~,\\
 \vert\phi_3\rangle&=&\tilde G_{-1}
\vert h\rangle_R{}~,{}~{}~{}~\vert\phi_4\rangle=
 \tilde G_{-1} \tilde G_0\vert h\rangle_R{}~,
\end{eqnarray}
and the Kac determinant
\begin{equation}
 \det({\cal M}^R_{17/16})={256\over81}(h+1/48-c/24)^2
 (h-h_{1,2})(h-h_{2,1})(h-h_{1,4})(h-h_{4,1}){}~.
\end{equation}

{}({\em d}) The $R^{(\pm)}$-modules, or the $[\chi^{3/2}_{3/2}]$ and
$[\chi^{1/2}_{1/2}]$ modules. $P_{1/2} (N)$ gives the number of states
at level $N$ in the $R^{(\pm)}$-modules built from
the highest weight states $\vert h;g^\pm \rangle$:
\begin{equation}
 G_0 \vert h;g^\pm \rangle = g^\pm \vert h;g^\pm \rangle{}~.
\end{equation}
With our normalization of $G_0$ in Eqn.(\ref{g^pm}),
\begin{equation}
 G_0=\left\{
 \begin{array}{ll}
 \tilde G_0 & {}~{}~{}~\mbox{if $c\leq 8$}{}~,\\
 i\tilde G_0 & {}~{}~{}~\mbox{otherwise}{}~.
 \end{array} \right.
\end{equation}
The eigenvalues $g^\pm$ are given by
\begin{equation}
 g^\pm={\sqrt{8-c}\over8\sqrt{6}}\pm\sqrt{h+1/48-c/24}{}~.
\end{equation}
The corresponding new determinant formulae read
\begin{equation}
 \det({\cal M}^{R^{(\pm)}}_N)=\alpha^{Q(R)} \prod_{r,s}
 [\alpha(g^\pm-g_{r,s})(g^\pm-g_{s,r})]^{P_{1/2} (N-{rs/4})}{}~,
\end{equation}
where $r,s$ are the same as in (\ref{kacdet-R}) and
\begin{equation}
 Q(R)=\sum_{n} P_{1/2} (N-n^2){}~.
\end{equation}
Here $n^2\leq N$.

{}The zeros of the
determinants are given by
\begin{equation}
 g_{r,s}={\sqrt{8-c}\over8\sqrt{6}}+
 {1\over\sqrt{96}}{\rm sgn}
 (j-1)\left[(r+s)\sqrt{2-c}+(r-s)\sqrt{8-c}\right]{}~,
\end{equation}
where $2j=(s-r){}~{\rm mod}{}~4$.

{}As an example, we consider the $R^{(\pm)}$-modules at the lowest
non-trivial level $N=9/16$. There is one state in each module,
\begin{equation}
 \vert\phi^\pm \rangle=\tilde G_{-{1\over2}}\vert h;g^\pm \rangle{}~.
\end{equation}
The new determinants can be calculated using Eqn.(\ref{LGcommutators})
and the GCR (\ref{GCR2})
\begin{equation}\label{newdet-9/16}
 \det({\cal M}^{R^{(\pm)}}_{9/16})={4\over3}\alpha(g^\pm -g_{1,2} )(g^\pm
 -g_{2,1}){}~.
\end{equation}

{}The Kac determinant (\ref{kacdet-R-9/16}) is related to the new
determinants (\ref{newdet-9/16}) via
\begin{equation}
 \det({\cal M}^R_{9/16})=\alpha(h+1/48-c/24)\det({\cal M}^{R^{(+)}}_{9/16})
 \det({\cal M}^{R^{(-)}}_{9/16}){}~.
\end{equation}
This factorization generalizes to all levels.

{}Some of the examples of the Kac and new determinants presented
in this section will be
useful later for the discussion of the null state structure of the
critical $K=4$ fractional superstring theory. The Kac spectrum of highest
weights $h_{r,s}$ for the $K=4$ case was first given in Ref \cite{FZft}.

%============================================================================
\widetext
\newpage
\section{Derivation of Kac and New Determinant Formulae}
\bigskip

{}In this section we derive the Kac and new determinant formulae for
an arbitrary (integer) level $K$ FSCA using the BRST operator \cite{ALT,CLT}
in a particular representation of
FSCAs via a non-interacting theory of the ${\bf Z}_K$ PF and a single
boson with the background charge. Since by definition the Kac and new
determinants for a given algebra are representation independent, our
derivation is valid for all the representations of FSCAs with modules
classified in section II.

{}In subsection A we deduce the highest
weights of the primary fields for which there are null states in the
modules. The Kac and new determinants vanish for these values of
$h$. In subsection B we derive the conformal dimensions of the null
states. The difference between these conformal dimensions and the highest
weights of the primary fields gives the lowest levels at which the null
states appear
in the corresponding modules. There we also derive the multiplicities of
the null states at
higher levels. This fixes the orders of the zeros of the Kac and new
determinants. In subsection C we point out the relation between the Kac
and new determinants, and via this relation derive the contributions in
the Kac determinants due to the fractional supercurrent zero mode.

\subsection{BRST Operators}
\medskip

{}We consider the representation of the level $K$ FSCA constructed by
turning on the background charge of the $\varphi$ boson in the $SU(2)_K$
WZW model discussed in section II. After the background charge is turned
on, the energy-momentum tensor of the $\varphi$ boson reads
\begin{equation}\label{energytensor}
 T_\varphi =-{1\over4}(\partial\varphi)^2+i{{\alpha}_0}{\partial}^2
 \varphi{}~,
\end{equation}
The total energy-momentum tensor of the theory is given by
\begin{equation}\label{Ttot}
 T=T_\varphi +T_{\rm PF}{}~,
\end{equation}
where $T_{\rm PF}$ is the energy-momentum tensor of the ${\bf Z}_K$ PF.
The total central charge is
\begin{equation}\label{centralcharge}
 c=c_0-24{\alpha_0^2}\leq c_0{}~.
\end{equation}

{}Once we turn on the background charge, the $SU(2)_K$ symmetry is broken.
However, the off-diagonal currents $J^\pm(z)$ (see Eqn.(\ref{Jcurrents}))
can be modified so that they remain the spin-1 screening currents:
\begin{equation}\label{screencurrent}
 S^\pm (z)=\phi^0_{\pm1}(z)\cdot\exp (i\alpha_\pm \varphi(z)){}~,
\end{equation}
where
\begin{equation}
 {\alpha}_\pm={\alpha}_0\pm\sqrt{{\alpha}_0^2+{1\over K}}{}~.
\end{equation}
The only remaining symmetry that survives the presence of the background
charge is the fractional superconformal symmetry.
The fractional supercurrent $G$ commutes with the screening charges
$S^{(\pm)}$, {\em i.e.},
\begin{equation}\label{SGOPE}
 S^\pm (z)G(w){}~\sim{}~{W(w)\over {(z{}~-{}~w)^2}}{}~+{}~{\rm reg}.{}~,
\end{equation}
where the single pole term is absent ($W(w)$ is some operator) and $G$ is
given by
\begin{equation}\label{Gcurrent}
 G={\sqrt{K}\over2}
 \left\{[\epsilon\partial\varphi-i{\alpha_0}(K+2)\partial{\epsilon}]
 +{iK({{\alpha}_+}-{{\alpha}_-})\over{K+4}}\eta\right\}{}~.
\end{equation}
If we take $\alpha_0$ to zero we will recover Eqn.(\ref{WZWGcurrent}).

{}The currents $T(z)$ and $G(z)$, defined in (\ref{Ttot}) and
(\ref{Gcurrent}), generate the level $K$
FSCA (\ref{FSCA}) for the values of the central charge $c\leq c_0$
\cite{AGT}.

{}Now we consider the BRST operators \cite{ALT,CLT}:
\begin{equation}\label{BRSToperators}
 Q^{(\pm)}_p{}~\equiv{}~\prod_{i=1}^p \oint{{\rm d}z_i\over2\pi
 i}{}~{S^\pm}(z_i){}~,
\end{equation}
where the $z_j$ integration contour is inside of the $z_i$ contour for
$j>i$, and all contours start and end at $z_1$. Since the screening
currents, $S^\pm (z)$, are dimension one operators and their OPE with
$G(z)$ is given by (\ref{SGOPE}), we conclude that $Q^{(\pm)}_p$
commute with $T(z)$ and $G(z)$:
\begin{equation}\label{QTGcommutators}
 \left[ Q^{(\pm)}_p,T(z)\right]=0{}~,{}~{}~{}~
\left[ Q^{(\pm)}_p,G(z)\right]=0{}~.
\end{equation}

{}From the requirement that the primary fields
of Eqn.(\ref{SU(2)primary}) remain primary
with respect to $G(z)$ and $T(z)$, we
find the following expression for the possible primary fields:
\begin{equation}\label{primary}
 \chi^j_{j,n,n^\prime} (z)=\phi^j_j (z)\cdot
 \exp(i\beta_{n,n^\prime}\varphi(z)){}~,
\end{equation}
where $\beta_{n,n^{\prime}}={1\over 2}(1{}~-{}~n){}~
\alpha_++{1\over 2}(1{}~-{}~n'){}~
\alpha_-$ and $n,n^\prime$ are integers satisfying the condition
\begin{equation}\label{condition}
 2j=(n^\prime -n ){}~{\rm mod}{}~K{}~.
\end{equation}
This condition follows from the requirement that the BRST operators
$Q^{(\pm)}_p$ be well-defined on the Fock space (There is no conceptual
difference between $Q^{(+)}_p$ and $Q^{(-)}_p$, so from now on we
concentrate only on
$Q^{(+)}_p$). The operation
$Q^{(+)}_p\vert\phi^j_j\cdot\exp[i\beta_{n,n^\prime}\varphi(0)]\rangle$
is well defined if and only if the outer $z_1$ contour closes.
By the standard method we have
\begin{eqnarray}\label{Qaction}
 &&\prod_{i=1}^p{\rm exp}[i\alpha_+ \varphi(z_i)]{}~\exp
 [i\beta_{n,n^{\prime}}\varphi(0)]=\nonumber\\
 &&\prod_{i<k}(z_i{}~-{}~z_k)^{2\alpha_+^2}
 {}~\prod_{i=1}^p{}~z_i^{2\alpha_+\cdot\beta_{n,n^\prime}}{}~:\exp [ i\alpha_+
 \sum^p_{i=1}\varphi(z_i)+i\beta_{n,n^\prime}\varphi(0)]:{}~,\\
 &&\Psi_1(z_1)...\Psi_1(z_p) \phi^j_j(0)=\prod_{i<k}{}~(z_i{}~-{}~z_k)^{-2/K}
 {}~\prod_{i=1}^p{}~z_i^{-2j/K}{}~{\phi^j_{j+p}}(0)+...{}~.
\end{eqnarray}
(The latter equation follows from the fusion rules
(\ref{fusionrules}) and the
conformal dimensions of the PF fields given in (\ref{pfdims})).
Then we make the change of variables $z_i \to z_1\cdot u_i{}~{\rm
for}{}~i=2,3...,p$ and demand the exponent of $z_1$ be an integer (so that
the $z_1$ contour closes). This gives $p=n$ and the above condition
(\ref{condition}).

{}The conformal weight of the
field $\chi^j_{j,n,n^\prime}$ is given by
\begin{equation}\label{delta}
 \Delta^{(j)}_{n,n^{\prime}}={{c-c_0}\over24}+{1\over96}\left((n+n^\prime)
 \sqrt{c_0-c}+(n-n^\prime)\sqrt{c_1-c}\right)^2+\Delta^j_j{}~.
\end{equation}

{}In the next subsection we show that for $n=r>0$ and $n^\prime =s>0$
there are null states in the modules built from the highest weight
states $\vert\chi^j_{j,r,s}\rangle$. Therefore, Eqn.(\ref{delta}) gives
the values of the highest weight $h$ at which the Kac and new
determinants vanish. Thus, for $n=r>0$ and $n^\prime =s>0$ Eqn.(\ref{delta})
gives Eqn.(\ref{nullstates}).

\subsection{Conformal Weights and Multiplicities of Null States}
\medskip

{}To deduce the conformal weights and multiplicities of the null states
appearing in the modules, we consider the BRST mapping of the primary
states onto the null states. The following diagram illustrates the BRST
mapping:
\bigskip
\bigskip
%\begin{figure}[p]
\begin{center}
\begin{picture}(150,135)(0,-135)
\put(0,0){$\chi^j_{-j,-r,-s}$}
\put(0,-60){$\chi^{\ell}_{-j,-r,-s}$}
\put(4,-10){\vector(0,-33){33}}
\put(4,-70){\vector(0,-33){33}}
\put(0,-120){$\chi^{j^\prime}_{-j,-r,-s}$}
\put(115,-60){\vector(-1,0){70}}
\put(115,-120){\vector(-1,0){70}}
\put(125,-60){$\chi^{\ell}_{-\ell,r,-s}$}
\put(125,-120){$\chi^{j^\prime}_{-\ell,r,-s}$}
\put(129,-70){\vector(0,-33){33}}
\put(66,-50){$Q^{(+)}_r$}
\put(66,-110){$Q^{(+)}_r$}
\end{picture}
\end{center}
%\caption{The BRST mapping}
%\end{figure}

{}The vertical axis measures the conformal dimension of the states: The
vertical arrows indicate the action of the $T$ and $G$ current modes.
The $T$ current has $j=m=0$ quantum numbers, while
$G$ has $j=1$ and $m=0$. Hence in the
vertical direction only the spin $j$ changes. The BRST operator
$Q^{(+)}_r$ has $j=0$ and $m=r$. Therefore, in
the horizontal direction only the magnetic quantum number $m$ varies. In
particular, the action of $Q^{(+)}_r$ on a field increases its $m$
quantum number by $r$ up to periodicity (\ref{automorphism}). In the
diagram, $\chi^j_{m,n,n^\prime} \equiv
\phi^j_m (0)\cdot\exp[i\beta_{n,n^\prime}\varphi(0)]\vert0\rangle$.

{}Consider the highest weight state
$\chi^\ell_{-\ell,r,-s}$ with $r,s>0$. According to Eqn.(\ref{condition})
$2\ell =(s+r){}~{\rm mod}{}~K$. The BRST operator $Q^{(+)}_r$ maps this state
onto the state $\chi^\ell_{_-j,-r,-s}$ that is a descendent of the
highest weight state $\chi^j_{-j,-r,-s}$:
\begin{equation}
 \chi^\ell_{-j,-r,-s}=Q^{(+)}_r\chi^\ell_{-\ell,r,-s}{}~.
\end{equation}
Here $-j=(-\ell+r){}~{\rm mod}{}~K=-{{(s-r)}\over2}{}~{\rm mod}{}~(K/2)$ in
agreement with the condition $2j=(s-r){}~{\rm mod}{}~K$ following from
(\ref{condition}).
Since the operator $Q^{(+)}_r$ commutes with $T$ and $G$,
the positive modes of $T$ and $G$ annihilate $\chi^\ell_{-j,-r,-s}$,
because they annihilate the primary state
$\chi^\ell_{-\ell,r,-s}$. This, in
particular, means that $\chi^\ell_{-j,-r,-s}$ is a primary and descendent
at once, hence it is null. The dimension of this null state is the same
as of $\chi^\ell_{-\ell,r,-s}$ and equals $\Delta^{(\ell)}_{r,-s}$.

{}The modules $[\chi^j_{-j,-r,-s}]$ and $[\chi^j_{j,r,s}]$
($2j=(s-r){}~{\rm mod}{}~K$), built from the highest weight states
$\chi^j_{-j,-r,-s}$ and $\chi^j_{j,r,s}$, are
dual. The duality means that each state in $[\chi^j_{-j,-r,-s}]$ has a
dual state in $[\chi^j_{j,r,s}]$ and structurally these are
identical. In particular, the highest weight state
$\chi^j_{j,r,s}$ is dual to $\chi^j_{-j,-r,-s}$ in the sense that the
two-point correlation function
$\langle\chi^j_{-j,-r,-s}\vert\chi^j_{j,r,s}\rangle=1$: The highest weights of
the states $\chi^j_{-j,-r,-s}$ and $\chi^j_{j,r,s}$ are equal. Thus, the
null state $\chi^\ell_{-j,-r,-s}$ appearing in
the module $[\chi^j_{-j,-r,-s}]$ is
dual to the null state $\chi^\ell_{j,r,s}$ of the module
$[\chi^j_{j,r,s}]$. These null states have the same conformal dimensions.
The level of the state $\chi^\ell_{j,r,s}$ above the highest
weight state $\chi^j_{j,r,s}$, {\em i.e.}, the lowest level in the module
$[\chi^j_{j,r,s}]$ at which this null state appears with multiplicity one, is
given by
\begin{equation}
 N^{r,s}_0=\Delta^{(\ell)}_{r,-s}-\Delta^{(j)}_{r,s}={rs\over
 K}+\Delta^\ell_\ell-\Delta^j_j{}~.
\end{equation}
The modified level of this state, that we defined in Eqn.(\ref{N}), is then
\begin{equation}
 N_{r,s}={rs\over K}+\Delta^\ell_\ell{}~.
\end{equation}

{}The descendents of the null state $\chi^\ell_{-j,-r,-s}$, which is
primary, are null states as well.  However, those are not primary.
There is a one-to-one correspondence
between the descendents of the primary state $\chi^\ell_{-\ell,r,-s}$ and
the descendents of the null state $\chi^\ell_{-j,-r,-s}$ manifested in
the BRST mapping. Thus, the number of the descendents of
$\chi^\ell_{-j,-r,-s}$ with spin
$j^\prime$ at some level $N$ in the module $[\chi^j_{-j,-r,-s}]$
is the same as the number of the descendents of
$\chi^\ell_{-\ell,r,-s}$ with the same spin $j^\prime$
at the level $\tilde N_0=N-N_{r,s}$ above the highest weight state
$\chi^\ell_{-\ell,r,-s}$ in the module $[\chi^\ell_{-\ell,-r,s}]$.
The modified level, defined in
Eqn.(\ref{N}), corresponding to
$\tilde N_0$ is then $\tilde N=\tilde
N_0+\Delta^\ell_\ell=N-rs/K$. The number of descendents of
$\chi^\ell_{-\ell,r,-s}$ with spin $j^\prime$ at the level $\tilde N$ is
given by ${P^{j^\prime}_{-\ell}}(\tilde
N)(\equiv{P^{j^\prime}_{\ell}}(\tilde N)$), where $P^j_m (N)$ are
defined in Eqn.(\ref{P}). Because of duality, the counting of descendents of
the null state $\chi^\ell_{j,r,s}$ in the module $[\chi^j_{j,r,s}]$ is
the same as the counting of descendents of the null state
$\chi^\ell_{-j,-r,-s}$ in the module $[\chi^j_{-j,-r,-s}]$. The
orders of the zeros of the Kac and new determinants are precisely
given by this counting.

{}Note that only the highest weight states in the BRST cohomology are the
true highest weight states \cite{CLT,Felder}. In this sense,
only the states $\chi^j_{j,r,s}$ and $\chi^j_{-j,-r,-s}$ with $r,s>0$ are the
highest weight states, whereas the state $\chi^\ell_{-\ell,r,-s}$ is not.
The highest weight states $\chi^j_{j,r,s}$ and $\chi^j_{-j,-r,-s}$ are
dual to each other, that is, if $\chi^j_{j,r,s}$ is a bra vector in the
Fock space, then $\chi^j_{-j,-r,-s}$ is its corresponding ket vector. The
$\varphi$ momenta of these states add up to $2\alpha_0$ to make up for the
presence of the background charge so that their two-point correlation
function is non-vanishing.

{}Now we turn to the Kac determinant formulae deriving whose explicit form
is our primary goal here. We are ready to write down the Kac
determinant formula (up to a positive normalization constant independent of
$h$ and $c$) for the
module $[\chi^0_0]$ of an arbitrary level $K$ FSCA. Since the fractional
supercurrent zero mode $G_0$ does not act on the highest weight state in
this module, the determinant of inner products of states is a polynomial in
$h$ at the lowest non-trivial level. Then by induction it
is a polynomial at all the higher levels. This immediately follows from the
counting of the null states derived using the BRST operator. Thus, we
arrive at (\ref{kacdet1}).

{}In other modules,
however, the zero mode $G_0$ acts on the highest weight states. As
pointed out in section II, the determinants of inner products of states
are polynomials in $h$ only for some modules, but not for all. In the
next subsection we resolve this issue and derive the contribution
of the zero mode $G_0$ into the Kac determinant via the relation between
the Kac and more fundamental new determinants.

\subsection{Zero Mode Contribution}
\medskip

{}In this subsection we derive the zero mode contribution.
This will complete our
derivation of the Kac determinant formulae for the $[\chi^{K/4}_{K/4}]$
and $[\chi^j_j
\oplus\chi^{K/2-j}_{K/2-j}]$ modules ($0<j<K/4$) and the new
determinant formulae for the $[\chi^j_j]$ modules ($0<j\not=K/4$).

{}Consider the module $[\chi^j_j]$, $0<j\not=K/4$. The form of the zero mode
algebra is fixed by the OPEs (\ref{FSCA}) and
the structure constant $\lambda(c)$ (\ref{lambda(c)})
\begin{equation}\label{ZeroMode}
 \left(G^2_0-2a^j \sqrt{c_1-c}{}~G_0 -(L_0 -{\tilde h^j})\right)
 \vert\chi^j_j\rangle=0{}~,
\end{equation}
where we choose the normalization ${\cal N}$ of the $G_0$ operator
\begin{equation}
 G(z)\chi^j_j(0)={\cal N}z^{-(K+4)/(K+2)}G_0 \chi^j_j(0)+...
\end{equation}
so that in
(\ref{ZeroMode}) there are only two parameters $a^j$
and ${\tilde h}^j$. The $G_0$ mode is diagonal with respect to the highest
weight state $\vert\chi^j_j\rangle$,
\begin{equation}
G_0 \vert\chi^j_j\rangle=g^j\vert\chi^j_j\rangle{}~,
\end{equation}
and Eqn.(\ref{ZeroMode}) becomes
\begin{equation}\label{ZeroMode1}
 ({g^j})^2-2a^j g^j\sqrt{c_1-c}-(h -{\tilde h^j})=0{}~,
\end{equation}
where $h$ is the highest weight of $\vert\chi^j_j\rangle$.

{}To fix the quantities $a^j$ and ${\tilde h}^j$, we consider the eigenvalues
$g_{r,s}$ of the $G_0$ operator corresponding to the highest weight states
$\vert\chi^j_{j,r,s}\rangle$ created by the vertex operators
(\ref{primary}) in the representation (\ref{Gcurrent}) of FSCA via the ${\bf
Z}_K$ PF and a single boson with background charge. We have
\begin{equation}
 G_0\vert\chi^j_{j,r,s}\rangle=g_{r,s} \vert\chi^j_{j,r,s}\rangle{}~.
\end{equation}
We determine the form of $g_{r,s}$ from the explicit representation of
the $G(z)$ current via (\ref{Gcurrent}) and the primary fields $\chi^j_{j,r,s}$
via (\ref{primary}):
\begin{equation}\label{g_rs0}
 g_{r,s}=u\sqrt{c_1-c}+vb_{r,s}{}~.
\end{equation}
Here we introduced $c$-independent quantities $u$ and $v$, and
\begin{equation}
 b_{r,s} \equiv{1\over\sqrt{96}}\left[(r+s)\sqrt{c_0-c}+(r-s)
 \sqrt{c_1-c}\right].
\end{equation}

{}The quantities (\ref{g_rs0}) must satisfy Eqn.(\ref{ZeroMode1}) with
$h=\Delta^{(j)}_{r,s}$, where $\Delta^{(j)}_{r,s}$ is the
conformal dimension of the primary field $\chi^j_{j,r,s}$ and is given
by Eqn.(\ref{delta}). This condition completely determines $a^j$ and
$\tilde h^j$:
\begin{equation}
 u=a^j={|K-4j|\over{\sqrt{24}(K+4)}}{}~,
\end{equation}
and
\begin{equation}\label{tilde-h}
 {\tilde h}^j={c\over24}-c{}~({a^j})^2=
{{c{}~(j+1)(K+2-2j)}\over{3 (K+4)^2}}{}~.
\end{equation}

For the $[\chi^j_j]$ module, $0<j\not=K/4$, we have $v={\rm sgn}(j-K/4)$ and
we arrive at Eqn.(\ref{g_rs1}). For the
$[\chi^{K/4}_{K/4}]$ module $v=\pm1$ for each pair $r,s$:
\begin{equation}
 g^{(1,2)}_{r,s} =\pm b_{r,s}{}~.
\end{equation}

{}Once we have derived $a^j$ and ${\tilde h}^j$, we can solve
Eqn.(\ref{ZeroMode1}):
\begin{equation}
 g^j=a^j \sqrt{c_1-c}+{\rm sgn}(j-K/4)\sqrt{h-h^j_0}{}~,
\end{equation}
where
\begin{equation}\label{h_0}
 h^j_0 = {{c-c_0}\over24}+\Delta^j_j ={c\over24}-{{(K-4j)^2}\over{8K(K+2)}}{}~.
\end{equation}

{}Note that if $j=K/4$, the coefficient $a^j$
vanishes and
\begin{equation}
 h^{K/4}_0 =\tilde h^{K/4}\equiv c/24{}~.
\end{equation}
Thus, in the module $[\chi^{K/4}_{K/4}]$
\begin{equation}
 g^\pm =\pm\sqrt{h-c/24}{}~.
\end{equation}
The determinants of inner products of states, corresponding to the
eigenvalues $g^\pm$, read
\begin{equation}
 \det({\cal M}^{(K/4)}_{(\pm)} (N))=\alpha^{Q(K/4)}
 \prod_{r,s}[(g^{\pm}-g^{(1)}_{r,s})(g^{\pm}-g^{(2)}_{r,s})]^{{P_\ell}
 (N-rs/K)}{}~,
\end{equation}
where $r,s,\ell,N$ and $Q(K/4)$ are the same as in Eqn.(\ref{kacdet30}).

{}Note that
$(g^{\pm}-g^{(1)}_{r,s})(g^{\pm}-g^{(2)}_{r,s})
=(g^{\pm})^2-(g^{(1)}_{r,s})^2=h-h_{r,s}$.
This means that the determinants
\begin{equation}
 \det({\cal M}^{(K/4)}_{(\pm)} (N))=\alpha^{Q(K/4)}
 \prod_{r,s}(h-h_{r,s})^{{P_\ell} (N-rs/K)}{}~
\end{equation}
are the Kac determinants since they are polynomials in the highest weight
$h$. If $G_0$ is not diagonal with respect to the highest weight state,
there is the additional zero mode
contribution in the Kac determinant. This contribution is given by
$(h-c/24)^{P_{K/4} (N)}$. Thus, we arrive at Eqn.(\ref{kacdet3}).

{}The new determinant for the module $[\chi^j_j]$ ($0<j\not=K/4$)
is a polynomial in the $G_0$ eigenvalue $g^j$ and is given by
Eqn.(\ref{kacdet-g}).

{}Consider the $[\chi^j_j\oplus\chi^{K/2-j}_{K/2-j}]$ module ($0<j<K/4$)
built from the highest weight state
\begin{equation}\label{h-g}
 \vert h\rangle=\gamma\vert h;g^j\rangle+\delta\vert h;g^{K/2-j}\rangle{}~,
\end{equation}
where $\gamma\delta\not=0$ and
\begin{equation}
 \langle h;g^j\vert h;g^{j^\prime} \rangle=\delta_{j,j^\prime}{}~.
\end{equation}

{}We label the operators creating the states at level $N$
by $V_i,{}~i=1,...,P_j (N)$, where $P_j (N)$ is the number of states at this
level. Thus, the states in the $[\chi^j_j]$ module have the form
\begin{equation}
 \vert\phi_i\rangle=V_i\vert h;g^j \rangle{}~.
\end{equation}
The new determinant reads
\begin{equation}
 \det({\cal M}^{(j)} (N))=
 \det(\langle\phi_i\vert\phi_k\rangle)=
 \det(\langle h;g^j\vert Z_{ik} \vert
 h;g^j\rangle){}~,{}~{}~{}~Z_{ik}\equiv V^\dagger_i V_k{}~.
\end{equation}

{}The Kac determinant for the $[\chi^j_j\oplus\chi^{K/2-j}_{K/2-j}]$
module is given by the $(2P_j (N))\times(2P_j (N))$ matrix
\begin{eqnarray}
 &&\det({\cal M}^{(j)}_0 (N))= \left| \begin{array}{lr}
   \langle h\vert Z_{ik}\vert h\rangle & \langle h\vert Z_{ik^\prime}
   {G_0} \vert h\rangle \\
   \langle h\vert {G^\dagger_0} Z_{{i^\prime}k}\vert h\rangle & \langle h\vert
   {G^\dagger_0} Z_{{i^\prime}k^\prime} {G_0}\vert h\rangle
 \end{array} \right| \nonumber\\
 &&=[|\gamma\delta|^2(g^+ -g^- )^2]^{P_j (N)} \det ({\cal
 M}^{(j)} (N)) \det ({\cal M}^{(K/2-j)} N))\\
 &&=[2|\gamma\delta|^2\alpha(h-h^j_0 )]^{P_j (N)} \det({\cal M}^{(j)}
(N))\det ({\cal M}^{(K/2-j)} N)){}~.\nonumber
\end{eqnarray}
This completes the derivation of the Kac and new determinant formulae.

{}Note, that if $h=h^j_0$, the state $G_0 \vert h\rangle$ becomes
null and the Kac determinant vanishes, whereas the new determinants
are non-zero.

%============================================================================
\widetext
\newpage
\section{Minimal Unitary Series In FSCA}
\bigskip

{}In this section we deduce the values of $c$ and $h$ for which FSCAs
can have unitary representations. Statistical
mechanical systems near the second order phase
transition are always expected to be described by an effective unitary field
theory with a local order parameter. For the remainder in this section we
will thus confine our attention to unitary theories. We analyze
unitary representations of FSCAs using the Kac determinants
(\ref{kacdet1}), (\ref{kacdet3}) and (\ref{kacdet2}).
If for a given module the Kac determinant is negative at any level, it
means that
there are negative-norm states at that level and the representation is
not unitary. If the determinant is greater than or equal zero, further
investigation is needed to determine whether or not the
representation is unitary. We sketch the non-unitarity proof for FSCAs.
Our proof is closely parallel to that for the Virasoro and superconformal
minimal series \cite{FQS1,FQS2}.

{}We consider an arbitrary level $K$ FSCA. In the region
$c_0<c<c_1$, all $h_{r,s}$ with $r\not=s$ have non-vanishing imaginary
parts, while
all $h_{r,r}<0$. For $c=c_1$ all zeros of the Kac determinant are real
and satisfy the following inequality
\begin{equation}
 h_{r,s}\leq-j(j+1)(K+4)/K(K+2)\leq0,{}~{}~{}~2j=(s-r){}~{\rm mod}{}~K{}~.
\end{equation}
This means that in the region $c_0<c\leq c_1,{}~h>0$ the Kac determinant is
non-vanishing and that all of the Kac matrix eigenvalues are
positive. Indeed, as $h\rightarrow\infty$, the matrix
becomes dominated by its diagonal elements that are strictly
positive. On the boundary $c=c_0$ we have
\begin{equation}
 h_{r,s}=h_{s,r}=(r-s)^2/4K+\Delta^j_j{}~.
\end{equation}
Therefore, for $c=c_0$ and $h>0$ the Kac determinant vanishes at the points
$h=n^2/4K+\Delta^j_j,{}~n\in{\bf N},$ but does not become negative. Thus,
the Kac determinant poses
no obstacle in principle to having unitary representations of FSCA in the
region $c_0\leq c\leq c_1,{}~h>0$. The only unitary
representation with $h=0$ is the trivial one with $c=0$, while there are
no unitary representations with $h<0$.

{}When $c>c_1$ (\ref{FSCA}) shows that the $G$ current is anti-hermitian and
the structure constant $\lambda(c)$ is imaginary. Therefore, all FSCAs
with $c>c_1$ are
necessarily non-unitary unless $c_{111}$, defined in section III, vanishes.
When $K=1$ or $2$ this is exactly the case and there are no
analogs of non-unitarity proof for $c>25$ representations of the Virasoro
algebra and $c>27/2$ representations of the superconformal algebra.

{}In the region $0<c<c_0,{}~h>0$
the Kac determinant is definitely negative at some level except
for the points $(c,h)$ lying on the vanishing curves $h=h_{r,s} (c)$
where the determinant becomes zero. Even on these curves, however, all
points, but the ones where they intersect, have ghosts. This discrete
set of intersection points, where unitary representations of the FSCA are
not excluded, occur at the following values of the central charge
\begin{equation}\label{unitarymodels1}
 c_p=c_0- {6K\over {p(p+K)}} = {3K\over{K+2}} {\left(1-
 {2(K+2)\over{p(p+K)}}\right)}{}~,{}~{}~{}~p=3,4,...
\end{equation}
($p=2$ is the trivial theory with $c=0$). For each $p$ the allowed values of
$h$ are given by
\begin{equation}\label{unitarymodels2}
 h_{r,s}={{[(p+K)r-ps]^2 -K^2}\over{4Kp(p+K)}} +\Delta^j_j{}~,
\end{equation}
where $1\leq r<p,{}~1\leq s<p+K$.

{}Thus, we see that the necessary conditions for unitary highest weight
representations of the level $K$ FSCA are either (for $K=1,2$ there is no
upper bound on the central charge)
\begin{equation}
 {3K\over{K+2}}\leq c\leq{3(K+4)^2 \over{K(K+2)}}{}~,{}~{}~{}~h\geq0{}~,
\end{equation}
or (\ref{unitarymodels1}) and (\ref{unitarymodels2}). That the latter
two conditions are also
sufficient, {\em i.e.}, that there indeed exist unitary representations
of FSCAs for these discrete values of
$c$ and $h$, was shown in Ref \cite{CLT} via the GKO coset space
construction \cite{GKO}. The fractional superconformal unitary minimal model
with the central charge ${c_p},{}~p\geq3,$ can be realized by the
$SU(2)_K\otimes SU(2)_{p-2}/SU(2)_{K+p-2}$ coset model. An
example of a statistical mechanical system with $K=4$ fractional
supersymmetry is the tricritical 3-state Potts model.

%============================================================================
\widetext
\newpage
\section{No-Ghost Theorem and Null State Structure\protect\\
of Fractional Superstring}
\bigskip

\subsection{No-Ghost Theorem for Subcritical $K=4$ FSS}
\medskip

{}In this subsection we prove the no-ghost theorem for the space-time
bosonic sector of
the subcritical spin-4/3 fractional superstring (FSS).
We generalize the Brower-Thorn proof of the
no-ghost theorem for the bosonic string \cite{Thorn} using the Kac
determinant formulae for the $K=4$ FSCA presented in section II. Our
discussion closely parallels that of Ref \cite{Thorn},
so we shall be brief.

{}The physical state conditions for the space-time bosonic sector of the
spin-4/3 FSS read
\begin{equation}\label{physical}
 L_n\vert{\rm Phys};q\rangle=G^{\pm}_{n-(1\mp q)/3}\vert{\rm Phys};q
 \rangle=0{}~,{}~{}~{}~
 L_0\vert{\rm Phys};q\rangle=v_q \vert{\rm Phys};q\rangle{}~,{}~{}~{}~n>0{}~,
\end{equation}
where $v_q$ is the intercept of the physical states $\vert{\rm Phys};q
\rangle$ with the ${\bf Z}_3$ charge $q$. Since
\begin{equation}\label{+-}
 \vert{\rm Phys};+1\rangle=G^-_0 \vert{\rm Phys};-1\rangle{}~,
\end{equation}
we have $v_{+1}=v_{-1}$.

{}Our primary interest in this section is the $c=5$ representation of the
spin-4/3
FSS realized by three free bosons and the $SO(2,1)_2$ WZW theory \cite{AT2}.
This representation
has three dimensional flat Minkowski space-time as its target space, {\em
i.e.}, $SO(2,1)$ global Lorentz symmetry, allowing for the particle
interpretation of its scattering amplitudes. The tree level
scattering of the physical states does not couple
them to the spurious states. In this section we show that the bosonic
physical spectrum of this theory is free of negative norm states. This,
in particular, means that the tree level scattering is unitary.

{}The Fock space of this string theory consists of the states
\begin{equation}\label{FSSFock}
 \vert\{\lambda\},\{\rho\},\chi\rangle{}~\equiv{}~
 (\alpha^0_{-{n_1}})^{\lambda_1}...(\alpha^0_{-{n_p}})^{\lambda_p}
 (\epsilon^{(\pm)0}_{-r_1})^{\rho_1}...(\epsilon^{(\pm)0}_{-r_s})^{\rho_s}
 \vert\chi;q\rangle{}~,
\end{equation}
where the ordering is the same as in (\ref{spanSU(2)}).
Here $\alpha^0_{-{n_i}}$, $i=1,...p$, are the time-like ($\mu=0$)
creation operators of the world-sheet boson $X^\mu (z)$, $\mu=0,1,2$,
while $\epsilon^0_{-{r_k}}$, $k=1,...,s$, are the time-like
creation operators of the spin-1/3 world-sheet field $\epsilon^\mu (z)$.
The state
$\vert\chi\rangle$ is created from the state $\exp[ik\cdot
X(0)]\vert0\rangle$ with the momentum $k^\mu$ by the
space-like creation operators of the fields $X^i (z)$ and $\epsilon^i (z)$,
$i=1,2$. Even though the space-like components of $\epsilon^\mu (z)$
couple to the time-like component, $\vert\chi\rangle$ is a positive
norm state since $SO(3)_2$ is free of
ghosts and the space-like components do not change when we rotate
$SO(3)_2$ to $SO(2,1)_2$ (the bosonic
space-like creation operators certainly do not spoil unitarity).
The time-like operators $\alpha^0_{-n}$ and $\epsilon^0_{-r}$ create
negative norm states, {\em i.e.}, ghosts, in the Fock space of the FSS.
The goal of the no-ghost theorem is to prove that there are no ghosts among the
physical states that satisfy the above physical state conditions.

{}The states (\ref{FSSFock}) are at the level
\begin{equation}\label{level}
 N_0=\sum_{i=1}^p {\lambda_i}{n_i} + \sum_{k=1}^s \rho_k r_k
\end{equation}
above the state $\vert\chi\rangle$. The metric tensor of these states is
given by
\begin{equation}\label{metric}
 T_{N_0}
 (\{\lambda^\prime\},\{\rho^\prime\};\{\lambda\},\{\rho\})=\delta_{\lambda_1
 ,\lambda^\prime_1}...\delta_{\lambda_p ,\lambda^\prime_p}
 \delta_{\rho_1 ,\rho^\prime_1}...\delta_{\rho_s ,\rho^\prime_s}
 (-1)^{\nu(p,s)}{}~,
\end{equation}
where
\begin{equation}
 \nu(p,s)=\sum_{i=1}^p \lambda_k + \sum_{k=1}^s \rho_k{}~.
\end{equation}

{}The Fock space $\cal F$ that consists of the states
(\ref{FSSFock}) can be decomposed into the subspace $\cal R$ of the primary
states $\vert h;q\rangle$, defined as
\begin{equation}
 L_0\vert h;q\rangle=h\vert h;q\rangle{}~,{}~{}~{}~L_n\vert h;q\rangle=
 G^{\pm}_{n-(1\mp q)/3}\vert h;q\rangle=0{}~,{}~{}~{}~n>0{}~,
\end{equation}
and the subspace of
spurious states $\cal S$, defined as the orthogonal complement
of $\cal R$ in $\cal F$. Thus, ${\cal F}={\cal R}\oplus{\cal S}$.
The subspace $\cal S$ is spanned by the states
\begin{equation}\label{spanspurious}
 L^{\lambda_1}_{-{n_1}}...L^{\lambda_p}_{-{n_p}}(G^\pm _{-r_1})^{\rho_1}...
 (G^\pm _{-r_s})^{\rho_s}\vert h;q\rangle
\end{equation}
(where the ordering is the same as in (\ref{FSSFock}) and at least one of
the numbers $\lambda_i$ or $\rho_k$ is non-zero) provided that
their Kac determinant is non-vanishing (the latter condition ensures that
these states are linearly independent). The physical states then are the
states with $h=v_q =N_0 +\Delta_{\chi}$, where $\Delta_{\chi}$ is the
$L_0$ eigenvalue of $\vert\chi;q\rangle$.

{}Consider the physical states with the intercept $v_q$ and the subspace
${\cal S}_q$ of the spurious state space $\cal S$ that consists of the
states with the $L_0$ eigenvalue $v_q$ and the ${\bf Z}_3$
charge $q$. According to the Kac determinant formulae for the $K=4$ FSCA, as
$h\rightarrow-\infty$, the metric tensor of the states in ${\cal S}_q$ at
level $N_0$ coincides with (\ref{metric}). If we now require the
condition
\begin{equation}\label{Thorn}
 \det({\cal M}_N)\not=0{}~{}~{}~{\rm for{}~all}{}~h<v_q -N_0
\end{equation}
be satisfied at any level $N_0 >0$, then
the counting of states in ${\cal S}_q$ at level $N_0$, given by
(\ref{level}), is the same as the counting of states
(\ref{FSSFock}) with the $L_0$ eigenvalue $v_q$. Moreover, the metric
of spurious states in ${\cal S}_q$ coincides with
(\ref{metric}) at any level since the Kac determinant is a
polynomial in $h$ and the condition (\ref{Thorn}) is satisfied. This, in
particular, means that the number of negative norm states in the Fock
space with the $L_0$ eigenvalue $v_q$ is the same as the
number of negative norm states in ${\cal S}_q$. The latter subspace
by definition consists only of spurious states, therefore, there
are no ghosts among the physical states. This establishes the no-ghost
theorem for the physical states with the intercept $v_q$.
In addition to the above condition
we should also check the level
zero for the absence of ghosts because the $G^\pm_0$ zero modes
may create negative-norm physical states.
Note that the above analysis can be applied to any representation of the
spin-4/3 FSS
with only one time-like direction. Eqn.(\ref{Thorn}) is then a
sufficient condition for the absence of ghosts in the physical spectrum.

{}The ${\bf Z}_3$ charge
$q$ is a quantum number conserved in the tree level scattering processes.
We refer to the physical states with
$q=\pm1$ as the $V$-sector, and the physical states with $q=0$
as the $T$-sector. The lowest lying V-sector state
is a massless vector particle, whereas its $T$-sector counterpart is a
tachyon.

{}The states that are both spurious and physical are the
so-called null states, {\em i.e.}, they have zero norm. The null states
in the $V$-sector come from the $S^{(\pm)}$- and $D^{(\pm)}$-submodules;
the null states in the $T$-sector are those
in the $S^{(0)}$- and $D^{(0)}$-submodules.

{}Now we turn to the ghost structure in the $V$-sector. If $c<8$ the only
real zeros of the Kac determinant relevant to the $V$-sector appear in
the $S^{(\pm)}$-submodules and are given by
\begin{equation}\label{hrr}
 h_{r,r}={{c-2}\over24}(1-r^2)>{1\over4}-{r^2\over4}{}~,
\end{equation}
where $r$ is an odd integer.
On the other hand, $N_0\geq N^{r,r}=r^2 /4+1/12$ and $v_{\pm1}
-N_0<v_{\pm1} -1/12-r^2 /4$. Therefore, taking into
account the inequality (\ref{hrr}) we conclude that in the $V$-sector the
condition (\ref{Thorn}) is always satisfied for intercepts $v_{\pm1}<1/3$ in
representations with $c<8$. However, in this sector the zero modes may
spoil unitarity. {}From (\ref{+-}) we find that if $\vert{\rm
Phys};-1\rangle$ is a positive-norm physical state, then $\vert{\rm
Phys};+1\rangle$ has the norm
\begin{equation}
 \langle{\rm Phys};+1\vert{\rm Phys};+1\rangle=(v-c/24)\langle{\rm Phys};
 -1\vert{\rm Phys};-1\rangle{}~,
\end{equation}
that is non-negative if and only if $v\geq c/24$. Thus, we established
the absence of ghosts in the $V$-sector for the following range of
$v\equiv v_{\pm1}$ and $c$:
\begin{equation}\label{V-sector}
 V-{\rm sector}:{}~{}~{}~c/24\leq v\leq1/3{}~,{}~{}~{}~c\leq8{}~.
\end{equation}
The case $v=1/3$ and/or $c=8$ was included by continuity.

{}Next we consider the $T$-sector. If $c<8$, the only real zeros of the Kac
determinant relevant to the $T$-sector appear in the $S^{(0)}$-submodule
and are given by (\ref{hrr}) with $r$ being an even integer. We
established the absence of ghosts in the $T$-sector for the following values of
$v\equiv v_0$ and $c$:
\begin{equation}\label{T-sector}
 T-{\rm sector}:{}~{}~{}~v\leq(10-c)/8{}~,{}~{}~{}~c\leq8{}~.
\end{equation}

{}For $c<8$ the condition (\ref{Thorn}) is automatically satisfied in
the $R$-sector, because none of the zeros of the Kac determinant are real
in this region. The zero mode contribution restricts the intercept to be
greater than $c/24-1/48$. Thus, we established the absence of ghosts in
the $R$-sector for the following values of $v_R$ and $c$:
\begin{equation}
 R-{\rm sector}:{}~{}~{}~v_R \geq c/24-1/48,{}~{}~{}~c\leq8{}~.
\end{equation}

\subsection{Null State Structure of Critical FSS}
\medskip

{}Any consistent string theory is expected to have extra sets of physical
null states at the critical central charge. It is yet unclear if
fractional superstrings exist as consistent string theories. Nonetheless,
the Kac determinant formulae for FSCAs allow us to examine the null state
structure of plausible critical FSS with an arbitrary level $K$
world-sheet fractional supersymmetry. In particular, we recover the well-known
results for the bosonic ($K=1$) and superstring ($K=2$) theories. We
analyze the null state structure of the spin-4/3 ({\em i.e.}, $K=4$) FSS
at the critical central charge and find extra sets of zero-norm physical
states.

{}First we review the discussion of Ref \cite{ALT}.
The zeros of the Kac determinants for the level $K$ FSCA are given by
\begin{equation}\label{kacdetzeroes}
 h_{r,s}={{c-c_0}\over24}+{1\over96}\left((r+s)
 \sqrt{c_0-c}+(r-s)\sqrt{c_1-c}\right)^2+\Delta^j_j{}~.
\end{equation}
Here $r,s\in{\bf N}$; $c_0=3K(K+2)$ and $c_1=24/K+c_0$;
$\Delta^j_j=j(K-2j)/K(K+2)$, where $2j=(s-r){}~{\rm mod}{}~K$.
The Kac determinant for the modules with
$j\not=0$ also vanishes at $h^j_0 =(c-c_0 )/24+\Delta^j_j$ (see section III).

{}The zero $h=h_{r,s}$ ({\em i.e.}, the null state) first appears at the level
\begin{equation}
 N^{r,s}_0={rs\over K}+\Delta^\ell_\ell-\Delta^j_j{}~,
\end{equation}
where $2\ell=(s+r){}~{\rm mod}{}~K$. The intersept of a physical null state
is given by
\begin{equation}\label{interceptv}
 v=h_{r,s}+N^{r,s}_0.
\end{equation}

{}Regardless of the central charge, there always is a null state at level
$N^{1,1}_0$ above the highest weight state $\vert
h_{1,1}\rangle=\vert0\rangle$. For $K>1$ the intercept of this state is
\begin{equation}\label{interceptV}
 v{}~={}~{2\over{K+2}}{}~;
\end{equation}
for $K=1$ we have $v=1$. This physical null state belongs to the sector
of the FSS theory that contains the massless vector particle (in open FSS)
or the graviton (in closed FSS).

{}The second set of physical null states that belong to the same sector
is built from the highest weight state $\vert h_{1,K+1}\rangle$. Their
intercept must be the same as that of Eqn.(\ref{interceptV}). On the
other hand, according to Eqn.(\ref{interceptv})
\begin{equation}
 v = h_{1,K+1}+N^{1,K+1}_0{}~.
\end{equation}
This set of constraints has a solution only at the critical central
charge \cite{ALT}
\begin{equation}\label{critical}
 c_{\rm critical}=c_1+c_0={6K\over{K+2}}+{24\over K}{}~.
\end{equation}

{}For the bosonic string theory $c_{\rm critical}=26$, $v=1$. For the
space-time bosonic sector of the superstring theory we also obtain the
well-known result: $c_{\rm critical}=15$, $v=1/2$. In the fermionic, or
Ramond sector of the superstring theory there also are two sets of
physical null states. First appears at level zero due to the vanishing of
the Kac determinant at $h=c/24$. This gives the Ramond sector intercept
$v_R=c/24=5/8$. The second set with the same
intercept $v_R$ appears at the level $N^{1,2}_0 =1$ above the highest
weight state $\vert h_{1,2}=-3/8\rangle$ ($c=15$).

{}Eqn.(\ref{critical}) gives $c_{\rm critical}=10$ for the $K=4$ FSS.
According to Eqn.(\ref{interceptV}) the intercept of the $V$-sector
physical states  is $v=1/3$.

{}In the $T$-sector at $c=10$ there also are two sets of
physical null states with the intercept
$v=1/3$. The first one occurs at level $N^{1,3}_0=2/3$
in the $D^{(0)}$-submodule built
from the highest weight state $\vert h_{1,3}=-1/3\rangle$. However, the
second set appears at a rather high level $N^{2,6}_0=3$
in the $S^{(0)}$-submodule. It is unclear if this will pose a
problem for unitarity.
The no-ghost theorem for the critical $K=4$ FSS is needed to
answer this question. In any case, it is likely that the $T$-sector (that
contains the tachyonic ground state) should be projected out
in any consistent string model.

{}There is another possibility if the
$T$- and $V$-sectors can have different intercepts. There are two
sets of physical null states in the
$T$-sector at $c=10$ and $v=0$. The first set appears at level $N^{2,2}_0=1$
in the $S^{(0)}$-submodule
built from the highest weight state $\vert h_{2,2}=-1\rangle$,
and the other one occurs at level $N^{1,7}_0=5/3$ in the $D^{(0)}$-submodule
built from the highest weight state $\vert h_{1,7}=-5/3\rangle$. The
ground state in this case is no longer tachyonic, but massless.

{}Extra sets of physical null states also occur in the $R$-sector
at $c=10$ and the intercept $v=3/8$. The first set
appears at level $N^{1,2}_0=1/2$ above
the highest weight state $\vert h_{1,2}=-1/8\rangle$, and the other one
occurs at level $N^{1,4}_0=1$ above the highest weight state
$\vert h_{1,4}=-5/8\rangle$.

{}The above analysis of the physical null state structure of the $K=4$ FSS
indicates that at least the $V$- and $R$-sectors can be expected to be free
of ghosts at the critical central charge.

\section{Conclusions}
\bigskip

{}In this paper we presented and derived the Kac and new determinant
formulae for an arbitrary integer level $K$ fractional superconformal algebra.
Thus, although complicated, the FSCAs can
be studied using tools developed in conformal field theories. Now we know the
Kac and new determinant formulae for infinitely many algebras, and only for
three of them these determinants can be explicitly calculated. These are the
(super)Virasoro and the spin-4/3 parafermion current algebras, for
which the (generalized) (anti)commutation relations are known. For the
rest of the FSCAs the generalized commutation relations have not yet been
written down because of the complications due to the non-abelian braiding
properties of these algebras.

{}Since all of the necessary tools for the rational level $K$ FSCAs have
been worked out \cite{ACT} (namely, the ${\bf Z}_{p/q}$ parafermion
theory, the rational level $K$ string functions and the BRST operators),
it is straightforward to generalize the Kac and new determinants to those
algebras.

\acknowledgments
\bigskip

{}It is a pleasure to thank Philip Argyres for useful discussions and
comments. This work was supported in part by the National Science
Foundation.

%============================================================================
\newpage
\widetext

\end{document}